\documentclass[10pt]{article}

\usepackage[english]{babel}
\usepackage[utf8x]{inputenc}
\usepackage{amsmath}
\usepackage{amsfonts}
\usepackage{graphicx}
\usepackage{color}
\usepackage{hyperref}

\usepackage[left=2.8cm,right=2.8cm,top=2.8cm,bottom=2.8cm]{geometry}

\newcommand\blfootnote[1]{%
  \begingroup
  \renewcommand\thefootnote{}\footnote{#1}%
  \addtocounter{footnote}{-1}%
  \endgroup
}

\newcommand\gn{\overset{\text{\scalebox{.5}{(0)}}}{g}{}}

\def\comma{\,,\,}
\def\triangledown{\nabla}
\def\tr{\text{tr}}
\newcommand{\tightoverset}[2]{%
  \mathop{#2}\limits^{\vbox to -.5ex{\kern-0.75ex\hbox{$#1$}\vss}}}

\def\cA{\mathcal{A}}
\def\cB{\mathcal{B}}
\def\cC{\mathcal{C}}
\def\cD{\mathcal{D}}
\def\cE{\mathcal{E}}

\def\cK{\mathcal{K}}

\def\cS{\mathcal{S}}
\def\cT{\mathcal{T}}

\numberwithin{equation}{section}

\begin{document}

\vspace*{-1.5cm}
\thispagestyle{empty}
\begin{flushright}
AEI-2014-036
\end{flushright}
\vspace*{2.5cm}
\begin{center}
{\Large
{\bf Metric- and frame-like higher-spin gauge theories in\\[2mm]
three dimensions}}
\vspace{2.5cm}

{\large Stefan Fredenhagen, Pan Kessel}
\blfootnote{{\tt E-mail: FirstName.LastName@aei.mpg.de}}

\vspace*{0.5cm}

Max-Planck-Institut f{\"u}r Gravitationsphysik\\
Albert-Einstein-Institut\\
Am M{\"u}hlenberg 1\\
14476 Golm, Germany\\
\vspace*{3cm}

{\bf Abstract}
\end{center}
We study the relation between the frame-like and metric-like
formulation of higher-spin gauge theories in three space-time
dimensions. We concentrate on the theory that is described by an
$SL(3)\times SL(3)$ Chern-Simons theory in the frame-like
formulation. The metric-like theory is obtained by eliminating the
generalised spin connection by its equation of motion, and by
expressing everything in terms of the metric and a spin-3 Fronsdal
field. 

We give an exact map between fields and gauge parameters in both
formulations. To work out the gauge transformations explicitly in
terms of metric-like variables, we have to make a perturbative
expansion in the spin-3 field. We describe an algorithm how to do this
systematically, and we work out the gauge transformations to cubic
order in the spin-3 field. We use these results to determine the gauge
algebra to this order, and explain why the commutator of two spin-3
transformations only closes on-shell.

\newpage

\tableofcontents


\section{Introduction}

Higher-spin gauge theories have gained a lot of attention in recent
years, in particular because of the proposed higher-spin AdS/CFT correspondence in
four and three dimensions (see~\cite{Giombi:2012ms,Gaberdiel:2012uj} for reviews).
Higher-spin gauge fields can either be described by extending the
vielbein formalism of gravity to higher-spins~\cite{Vasiliev:1980as}, or by extending the
metric formulation~\cite{Fronsdal:1978rb}. Although the metric-like description might be the
more intuitive ansatz, because one needs less auxiliary fields, it is
the frame-like formulation that allowed Vasiliev to construct a
consistent non-linear theory of interacting higher-spin gauge
fields~\cite{Vasiliev:1990en,Vasiliev:2003ev}. In the metric-like formulation, on the other hand, one only
knows how to construct interactions in a perturbative expansion, e.g.\
one has obtained a classification of consistent cubic
terms~\cite{Bengtsson:1983pd,Metsaev:2005ar,Bekaert:2010hp,Manvelyan:2010jr,Sagnotti:2010at,Joung:2011ww,Joung:2012fv,Joung:2012hz,Joung:2013nma}.

It would be desirable to understand the theory also in the metric-like
formulation. In particular one would hope that one could get a better
geometric understanding of the higher-spin gauge symmetry as
generalised diffeomorphisms. This might also improve our understanding
of particular solutions of higher-spin theories like higher-spin
analogues of black holes~\cite{Gutperle:2011kf,Ammon:2011nk,Castro:2011fm,Banados:2012ue,Bunster:2014mua}. In~\cite{Campoleoni:2012hp} it was shown how one could use
the Wald formula in a metric-like higher-spin formulation to compute
the entropy of higher-spin black holes (for other approaches see e.g.\
\cite{Gutperle:2011kf,Kraus:2011ds,Gaberdiel:2012yb,Perez:2012cf,Perez:2013xi,deBoer:2013gz,Kraus:2013esi,Ammon:2013hba,deBoer:2013vca,Compere:2013nba,Datta:2014ska}).

Higher-spin gauge theories in three dimensions are considerably
simpler than in higher dimensions, because they do not contain propagating degrees of freedom
and can be written as a Chern-Simons
theory~\cite{Blencowe:1988gj,Bergshoeff:1989ns}. Also, in contrast to
higher dimensions, it is possible to truncate the tower of
typically infinitely many higher-spin gauge fields to a finite
selection -- the simplest theory only contains gravity and one spin-3
field. In this case the generalised vielbein $e=e_{\mu}dx^{\mu}$ and
the generalised spin connection take values in the Lie algebra
$sl(3,\mathbb{R})$,
\begin{equation}
e_{\mu} = e_{\mu}^{\cA} \,J_{\cA} \quad ,\quad \omega_{\mu} =
\omega_{\mu}^{\cA} \,J_{\cA} \ ,
\end{equation}
where $J_{\cA}$ form a basis of $sl(3,\mathbb{R})$,
\begin{equation}
[J_{\cA},J_{\cB}] = f_{\cA\cB}{}^{\cC} \, J_{\cC} \ .
\end{equation}
The gauge sector of this theory is described by the action
\begin{equation}
\label{eq:action}
S = \frac{1}{16\pi G} \int \tr \left( e \wedge R + \frac{1}{3l^2} e
\wedge e \wedge e \right)\ ,
\end{equation}
where 
\begin{equation}\label{eq:curvatureTensor}
R=d\omega +\omega\wedge\omega \quad \Leftrightarrow \quad R^\mathcal{A} = d \omega^\mathcal{A} +
\frac{1}{2}\,f^{\mathcal{A}}{}_{\mathcal{B}\mathcal{C}} \,\omega^\mathcal{B} \wedge
\omega^\mathcal{C}
\end{equation}
is the curvature of
the generalised spin connection, $G$ is the gravitational constant and
$\tr$ is the trace in the fundamental representation of $sl(3, \mathbb{R})$.
The parameter $l$ is related to the cosmological constant -- a
real and positive $l$ coincides with the radius of the AdS
solution. This action can be rewritten as a Chern-Simons theory whose
gauge group depends on the cosmological constant: e.g.\ for a negative
constant (positive $l^{2}$) the gauge group is 
$SL(3,\mathbb{R})\times SL(3,\mathbb{R})$.

The frame-like formulation being so simple, it is tempting to try to
reformulate it in terms of metric-like fields. First one has to
eliminate the spin connection by its equation of motion,
\begin{equation}
\label{eq:torsionConstraint}  
D_{[\mu } e_{\nu ]} = 0 \ ,
\end{equation}
where $D_{\mu}$ is the covariant derivative including spin connection
and the Levi-Civita Christoffel symbols,
\begin{equation}
\label{eq:covariantDer}
D_{\mu} e_{\nu}^{\cA} = \partial_{\mu}e_{\nu}^{\cA} +
f^{\cA}{}_{\cB\cC}\, \omega^{\cB}_{\mu}\, e_{\nu}^{\cC} -
\Gamma^{\lambda}{}_{\mu\nu} \,e_{\lambda}^{\cA} \ .
\end{equation}
Then one has to express everything in terms of metric-like fields,
which have to be expressions in the vielbeins where all $sl(3)$
indices are contracted with invariant tensors. In~\cite{Campoleoni:2010zq} it was proposed
to define the metric and the spin-3 field as
\begin{equation}
\label{eq:metric}
g_{\mu \nu} = \kappa_{\mathcal{A} \mathcal{B}} \,e^\mathcal{A}_\mu \,e^\mathcal{B}_\nu
\end{equation}
and
\begin{equation}
\label{eq:spin3Field}
\phi_{\mu \nu \rho} = \frac{1}{3!} \,
d_{\mathcal{A} \mathcal{B}\mathcal{C}}\, e^\mathcal{A}_\mu
\,e^\mathcal{B}_\nu \,e^\mathcal{C}_\rho \ .
\end{equation}
The remaining task is then to rewrite the action (after
eliminating $\omega$) in terms of these fields.

Because the vielbein is not invertible (it is not a square matrix),
this is rather complicated. In~\cite{Campoleoni:2012hp} the action was worked out
to quadratic order in the spin-3 field by making a general ansatz and
then demanding that explicit solutions of the frame-like theory should map to
solutions in the metric-like theory. We will follow here a different
approach which was started also in~\cite{Campoleoni:2012hp}.\footnote{For an alternative ansatz for a metric-like description see~\cite{Fujisawa:2012dk,Fujisawa:2013lua}.}

Instead of considering the action and its solutions we concentrate on
the gauge transformations. We formulate an exact map of the gauge
parameters in the frame- and the metric-like formulation
(section~2). Notice that one always has the freedom to reparameterise
the gauge transformations, therefore this map is not unique. We then
formulate an algorithm that can be used to map any given contraction
of frame-like quantities to metric-like quantities in a perturbative
expansion in the spin-3 field (section~3). We use this algorithm to
explicitly compute the gauge transformations in the metric-like theory
to cubic order. We are then in the position to compute commutators of
these transformations to better understand the gauge algebra in the
metric-like theory (section~4). We also discuss there why the
commutator of two spin-3 transformations only closes on-shell.
\smallskip

Our work clarifies a few issues that were left unanswered in~\cite{Campoleoni:2012hp}.
First of all we could show that the perturbative ansatz for the map between gauge
parameters in~\cite{Campoleoni:2012hp} can be used as an exact map without any
corrections, such that one has an exact dictionary of fields and
gauge parameters. Secondly we can explain why the gauge
transformations only close on-shell in the metric-like theory (whereas
they close off-shell on the frame-like side). Thirdly we worked out a
systematic approach to obtain the explicit expressions on the
metric-like side that does not require the use of specific solutions
of the theory. Last but not least we worked out the gauge
transformations and gauge algebra to one order higher than in~\cite{Campoleoni:2012hp} in
the hope to understand the metric-like theory better.

The expressions that we obtain for the gauge transformations to the
order we consider are already quite large, they fill two pages of the
appendix. In principle one could now go on and determine the
corresponding action (which is a fairly easy task if one uses a
powerful computer algebra program), and the result will be of similar
size. We have not found any pattern in our expressions that could help
to organise them -- but without such a pattern it does not make sense
to work out the metric-like theory to even higher orders. On the other
hand, one might hope that there is a clever redefinition of fields and
gauge parameters which makes the theory more manageable.


\section{Relating frame- and metric-like gauge transformations}

In this section we relate the gauge transformations in the
frame- and in the metric-like description. In the frame-like theory
there are the two types of gauge transformations; the generalised
local Lorentz transformations,
\begin{align}
\delta^{L}_{\Lambda} e_{\mu} &= [\Lambda ,e_{\mu}] \\
\delta^{L}_{\Lambda} \omega_{\mu} & = D_{\mu} \Lambda \ , 
\end{align}
and the generalised local translations,
\begin{align}
\label{eq:genLocalTranslations}
\delta_{\Xi} e_{\mu} & = D_{\mu} \Xi \\
\delta_{\Xi} \omega_{\mu} & = \frac{1}{l^2} \left[ e_\mu, \Xi \right] \ .
\end{align}
The local Lorentz transformations act trivially on all metric-like
fields built from the vielbeins $e_{\mu}$. The generalised local
translations, on the other hand, induce non-trivial transformations on
them, and they can be interpreted as diffeomorphisms and
higher-spin generalisations thereof.

Let us first consider pure diffeomorphisms. It is well-known (see
e.g.\ \cite{Witten:1988hc}) that a
generalised translation, where the parameter $\Xi$ is of the form
\begin{equation}\label{purediffeo}
\Xi^{\cA} = e^{\cA}_{\mu}\, \xi^{\mu} \ ,
\end{equation}
induces a diffeomorphism generated by the vector-field $\xi^{\mu}$
(up to a local Lorentz rotation) if one imposes the torsion
constraint~\eqref{eq:torsionConstraint}. The action of such a diffeomorphism (spin-2 gauge transformation) on any metric-like field $\phi$ built from the vielbeins $e$ is given by
\begin{equation}
\delta^{(2)}_{\xi}\phi  =  \mathcal{L}_{\xi}\phi  \ ,
\end{equation}
where $\mathcal{L}_{\xi}$ denotes the Lie derivative.

For the higher-spin transformations we do not know how they
act in general, but only in the linearised theory where they should
reproduce the transformations of free Fronsdal fields \cite{Fronsdal:1978rb}. The
spin-3 transformation should act as
\begin{align}
\delta^{(3)}_{\xi} g_{\mu\nu} & = 0 + \dotsb \\
\delta^{(3)}_{\xi} \phi_{\mu\nu\rho} & = \nabla_{\!(\mu} \Big(\xi_{\nu\rho)} -
\frac{1}{3}g_{\nu\rho)}\xi_{\lambda}{}^{\lambda}\Big) + \dotsb  \ ,  
\end{align}
where $\xi^{\mu\nu}$ is a symmetric tensor that labels the spin-3
gauge transformations, and the dots indicate terms that are at least
linear in the spin-3 field. The covariant derivative $\nabla_{\!\mu}$
is defined with respect to the Levi-Civita connection.

We combine the gauge parameters for spin-2 and spin-3 transformations
into a single object $\xi=(\xi^{\mu},\xi^{\nu\rho})$. We are looking for a map
\begin{equation}
\xi= (\xi^{\mu},\xi^{\nu\rho}) \mapsto \Xi(\xi) \ ,
\end{equation}
such that 
\begin{equation}
\delta_{\Xi(\xi)} \phi =
\delta^{(2)}_{\xi} \phi + \delta^{(3)}_{\xi}\phi \ . 
\end{equation}
Note that such a map is not unique, even if we have fixed the
expression of the metric-like fields in terms of frame-like ones such
that no field redefinitions are possible. We can still redefine the
higher-spin gauge parameters by terms that are at least linear in the
higher-spin fields, such that the linearised gauge transformations are
untouched. In the following we will construct one such map that is
valid at all orders in the spin-3 field.

\subsection{A proposal for the map}
\label{sec:proposalForTheMap}

The map $\xi\mapsto\Xi(\xi)$ is linear, so we can write it as
\begin{equation}
\label{eq:map}
\Xi^{\cA}(\xi) = \cS^{\cA}{}_{\mu}\, \xi^{\mu} +
\cS^{\cA}{}_{\nu\rho}\, \xi^{\nu\rho} \ ,
\end{equation}
with possibly field-dependent matrices $\cS$. The implementation of pure diffeomorphisms is
given by~\eqref{purediffeo}, this fixes the coefficients
$\cS^{\cA}{}_{\mu}$ to
\begin{equation}
\cS^{\cA}{}_{\mu} = e^{\cA}_{\mu} \ .
\end{equation}
An arbitrary frame-like gauge transformation $\Xi^{\cA}$ will induce
both a diffeomorphism and a spin-3 transformation, therefore there
will be projections $P$ and $(\mathbf{1}-P)$ such that $P\Xi$ induces
a pure diffeomorphism, and $(\mathbf{1}-P)\Xi$ a pure spin-3
transformation. Instead of fixing $\cS^{\cA}{}_{\nu\rho}$ directly, we
will rather first attempt to fix the projection $P$. It should project
an arbitrary gauge transformation to a pure diffeomorphism, therefore
we demand that
\begin{equation}
\text{for every}\ \Xi^{\cA}\ \text{there is a}\ \xi^{\mu}\ \text{such
that} \ P^{\cA}{}_{\cB}\, \Xi^{\cB} = \cS^{\cA}{}_{\mu}\, \xi^{\mu} \ ,
\end{equation}
and $P^{2}=P$. A natural requirement for the projector is that it is
orthogonal w.r.t.\ the Killing form, in other words that
\begin{equation}
P^{\cA\cB} = P^{\cB\cA} \ ,
\end{equation}
where we have raised the indices with the Killing form. This then
fixes the projector uniquely to be
\begin{equation}\label{projector}
P^{\cA\cB} = e^{\cA}_{\mu}\,g^{\mu\nu}\,e^{\cB}_{\nu} \ .
\end{equation}
Indeed we can easily check that 
\begin{align}
P^{\cA}{}_{\cB}\,P^{\cB}{}_{\cC} &=
e^{\cA}_{\mu}\,g^{\mu\nu}\,e^{\cD}_{\nu} \, \kappa_{\cB\cD}\,
 e^{\cB}_{\rho}\,g^{\rho\sigma}\,e^{\cE}_{\sigma}\, \kappa_{\cE\cC} \\
&= e^{\cA}_{\mu}\,g^{\mu\nu}\,g_{\nu\rho} \, g^{\rho\sigma}\,e^{\cE}_{\sigma}\, \kappa_{\cE\cC} \\
&= e^{\cA}_{\mu}\,g^{\mu\sigma}\,e^{\cE}_{\sigma}\, \kappa_{\cE\cC} =
P^{\cA}{}_{\cC} \ ,
\end{align}
where we used the definition of the metric~\eqref{eq:metric} to go to
the second line.
Furthermore, for an arbitrary $\Xi^{\cA}$ we have
\begin{equation}
P^{\cA}{}_{\cB}\,\Xi^{\cB} =
e^{\cA}_{\mu} \Big(
g^{\mu\nu}\,e^{\cC}_{\nu}\,\kappa_{\cC\cB}\,\Xi^{\cB}\Big) \ ,
\end{equation}
therefore $P$ indeed projects onto pure diffeomorphisms (where we
interpret the term in the parentheses as the corresponding vector
field). Notice that defining $P$ to be an orthogonal projector was a
choice we made, but we will see in the next section that by redefining
the gauge parameters it is always possible to bring the projector to
the form above.

Having fixed $P$ we can now look for an $\cS^{\cA}{}_{\nu\rho}$ that
satisfies
\begin{equation}
P^{\cA}{}_{\cB}\, \cS^{\cB}{}_{\nu\rho} = 0 \ . 
\end{equation}
In addition we want that $\cS^{\cA}{}_{\nu\rho}$ coincides with the free
field expression when we set the higher-spin fields to zero, i.e.\ 
\begin{equation}\label{linearisedgaugetrafo}
\cS^{\cA}{}_{\nu\rho} = 3\,d^{\cA}{}_{bc}\, e^{b}_{\nu}\,e^{c}_{\rho} +
\dotsb 
\end{equation}
A natural ansatz for a covariant expression that is consistent with
the linearisation and with the projection $P$ is then 
\begin{equation}
\cS^{\cA}{}_{\nu\rho} = \big(\delta^{\cA}_{\cD}- P^{\cA}_{\cD}\big) 3\,
d^{\cD}{}_{\cB\cC}\,
e^{\cB}_{\nu}\,e^{\cC}_{\rho} + \dotsb 
\end{equation}
The linearised gauge transformation does only depend on the
traceless part of the gauge parameter $\xi^{\mu\nu}$. If we want this
property to hold also at the non-linear level, we have to add the
projection to the traceless part,
\begin{equation}
\xi^{\mu\sigma} \mapsto \left(\delta^{\mu}_{\nu}\,\delta^{\sigma}_{\rho} -
\frac{1}{3}g^{\mu\sigma}\,g_{\nu\rho} \right) \, \xi^{\nu\rho} \ .
\end{equation}
Our final ansatz for $\cS^{\cA}{}_{\nu\rho}$ then reads
\begin{align}
\cS^{\cA}{}_{\nu\rho} &= \big(\delta^{\cA}_{\cD}- P^{\cA}_{\cD}\big) 3\,
d^{\cD}{}_{\cB\cC}\,
e^{\cB}_{\mu}\,e^{\cC}_{\sigma} \left(\delta^{\mu}_{\nu}\,\delta^{\sigma}_{\rho} -
\frac{1}{3}g^{\mu\sigma}\,g_{\nu\rho} \right)\\
&=  3\Big( d^{\cA}{}_{\cB\cC}\,e^{\cB}_{\mu}\,e^{\cC}_{\sigma} 
-6\,e^{\cA}_{\kappa}\,g^{\kappa\lambda}\,
\phi_{\lambda\mu\sigma}\Big)\left(\delta^{\mu}_{\nu}\,\delta^{\sigma}_{\rho} -
\frac{1}{3}g^{\mu\sigma}\,g_{\nu\rho} \right) \label{proposalforSAnurho}\ ,
\end{align}
where we used our definition for the spin-3 field $\phi$
in~\eqref{eq:spin3Field}.

To summarise we propose the following map for the gauge parameters,
\begin{align}
\Xi^{\cA} & = \cS^{\cA}{}_{\mu}\, \xi^{\mu} +
\cS^{\cA}{}_{\nu\rho}\,\xi^{\nu\rho} \\
&= e^{\cA}_{\mu}\, \xi^{\mu} + 
3\, \big(\delta^{\cA}_{\cD}- P^{\cA}_{\cD}\big) 
d^{\cD}{}_{\cB\cC}\,
e^{\cB}_{\mu}\,e^{\cC}_{\sigma} 
\left(\delta^{\mu}_{\nu}\,\delta^{\sigma}_{\rho} -
\frac{1}{3}g^{\mu\sigma}\,g_{\nu\rho} \right)
\xi^{\nu\rho} \ .
\label{proposedmap}
\end{align}
In the following section we will argue that this is a consistent
choice to all orders in the higher-spin field.

\subsection{The proposed map is exact}

Our goal is to obtain an exact map between the gauge parameters on
the metric-like side and on the frame-like side,
\begin{equation}\label{generalframe-metric-likemap}
\Xi^{\cA} = \cS^{\cA}{}_{M}\, \xi^{M} \ .
\end{equation} 
Here, $M$ is a collective label for the metric-like labels, e.g.\ in
the $sl(3)$ case $\{M\}=\{\mu,(\nu\rho)\}$, where $(\nu\rho)$ denote
symmetric pairs of space-time labels without any trace
constraints. The matrix $\cS$ is then not a square matrix, and it can
depend on the fields and on the vielbein. In the last section we have
made a proposal for such a map (see~\eqref{proposedmap}). In this
section we want to show that there will always be a redefinition of
the gauge parameters such that the proposal~\eqref{proposedmap} provides
the exact map.

Given a frame-like gauge parameter $\Xi^{\cA}$ we may ask what the
corresponding diffeomorphism and spin-3 transformation are that it
induces, in other words we want to have an inverse relation of the form
\begin{equation}
\xi^{M} = \cT^{M}{}_{\cA}\, \Xi^{\cA} \ ,
\end{equation}
such that
\begin{equation}
\cS^{\cA}{}_{M}\, \cT^{M}{}_{\cB} = \delta^{\cA}_{\cB} \quad , \quad
\cT^{M}{}_{\cA}\, \cS^{\cA}{}_{N} = \cK^{M}{}_{N} = \begin{pmatrix}
\delta^{\mu}_{\nu} & 0\\
0 & \cK^{\mu_{1}\mu_{2}}_{\nu_{1}\nu_{2}}
\end{pmatrix} \ . 
\end{equation}
Here, $\cK$ is a projector: not all components of $\xi^{\mu\nu}$
give rise to independent gauge transformations, and those $\xi$ that
are annihilated by $\cK$ do not contribute. Therefore $\cK$ projects
$\xi^{\mu\nu}$ to the part that contributes to non-trivial gauge
transformations. In the linearised approximation, $\cK$ projects onto
traceless tensors, in the full theory $\cK$ could act differently.
Of course this structure generalises straightforwardly to the
situation with more higher-spin fields.

Such a map between frame- and metric-like gauge parameters, specified
by $\cS$ and $\cT$, is not unique, because we can redefine the gauge
parameter on the metric-like side. Suppose we are given $\cS$ and
$\cT$, and the associated projector $\cK$. Then we can parameterise
$\xi^{M}$ by a new gauge parameter $\tilde{\xi}{}^{M}$,
\begin{equation}
\xi^{M} = \Phi^{M}{}_{N}\, \tilde{\xi}{}^{N} \ ,
\end{equation}
with a possibly field-dependent matrix $\Phi$. It does not need to be
invertible, but we want that $\tilde{\xi}$ still parametrises the
full set of gauge transformations. Therefore we request that
\begin{equation}
\text{Im} \big(\cK \Phi\big) = \text{Im}\big(\cK \big) \ .
\end{equation}
Then there is a map $\Psi$ in the opposite direction,
\begin{equation}
\tilde{\xi}{}^{M} = \Psi^{M}{}_{N}\, \xi^{N} \ ,
\end{equation}
such that 
\begin{equation}\label{KPhiPsi}
\cK \Phi \Psi = \cK \ .
\end{equation}
$\Psi$ acts as an inverse after projection by $\cK$.

With such a redefinition of the gauge parameters, the
map~\eqref{generalframe-metric-likemap} 
between frame-like and metric-like gauge parameters is changed into
\begin{equation}
\Xi^{\cA} = \tilde{\cS}{}^{\cA}{}_{M}\, \tilde{\xi}{}^{M}\quad ,\quad
\tilde{\cS} = \cS \Phi \ .
\end{equation}
Similarly we can introduce a new inverse map $\tilde{\cT}$,
\begin{equation}
\tilde{\xi}{}^{M} = \tilde{\cT}{}^{M}{}_{\cA}\, \Xi^{\cA}\quad ,\quad 
\tilde{\cT} = \Psi \cT \ , 
\end{equation}
such that
\begin{equation}
\tilde{\cS} \tilde{\cT} = \mathbf{1}\quad ,\quad 
\tilde{\cT} \tilde{\cS} = \Psi \cK \Phi =:\tilde{\cK}\ .
\end{equation}
Notice that the prescription is symmetric in the sense that we can also
view $\xi^{M}$ to provide a new parameterisation of
$\tilde{\xi}{}^{M}$ via $\Psi$, and the maps $\Phi$ and $\Psi$ satisfy
\begin{equation}
\tilde{\cK} \Psi \Phi = \tilde{\cK} \ ,
\end{equation}
in analogy to~\eqref{KPhiPsi}.
\medskip

Let us now apply this general discussion to the situation we are
interested in. We assume that there is an exact map relating the gauge
parameters as above with corresponding matrices $\cS$ and $\cT$ which
are a priori unknown. We then show that there is a redefinition of gauge
parameters such that the transformed $\cS$ coincides with our
proposal.

We already know how a pure diffeomorphism is implemented on the
frame-like side, therefore $\cS^{\cA}{}_{\mu}=e^{\cA}_{\mu}$ is fixed, and should
not be altered by a reparameterisation. We can then restrict to
matrices $\Phi$ and $\Psi$ of the form
\begin{equation}
\Phi =\begin{pmatrix}
\mathbf{1} & *\\
0 & *
\end{pmatrix} \quad ,\quad 
\Psi =\begin{pmatrix}
\mathbf{1} & *\\
0 & *
\end{pmatrix} \ .
\end{equation}
Suppose now that we have found a $\hat{\cT}{}^{\mu}{}_{\cA}$ such that 
\begin{equation}
\hat{\cT}{}^{\mu}{}_{\cA}\, \cS^{\cA}{}_{\nu} = \delta^{\mu}_{\nu} \ .
\end{equation}
In our case this will be given by 
\begin{equation}\label{hatT}
\hat{\cT}{}^{\mu}{}_{\cA} = g^{\mu\nu}\, e^{\cB}_{\nu}\,
\kappa_{\cB\cA} \ ,
\end{equation}
such that we recover our projector on diffeomorphisms
(see~\eqref{projector}) as
\begin{equation}
\cS^{\cA}{}_{\mu}\,\hat{\cT}{}^{\mu}{}_{\cB} = P^{\cA}{}_{\cB} \ .
\end{equation}
Then we set 
\begin{equation}
\Psi = \begin{pmatrix}
\mathbf{1} & \Psi^{\mu}{}_{\rho\sigma} \\
0 & \mathbf{1}
\end{pmatrix}\quad ,\quad 
\Psi^{\mu}{}_{\rho\sigma} = \big(\hat{\cT}{}^{\mu}{}_{\cA} -
\cT^{\mu}{}_{\cA} \big) \cS^{\cA}{}_{\rho\sigma} \ .
\end{equation}
With this transformation one finds 
\begin{align}
\tilde{\cT}{}^{\mu}{}_{\cA} &= \cT^{\mu}{}_{\cA} +
\Psi^{\mu}{}_{\rho\sigma}\, \cT^{\rho\sigma}{}_{\cA}\\
&= \cT^{\mu}{}_{\cA} +\big(\hat{\cT}{}^{\mu}{}_{\cB} -
\cT^{\mu}{}_{\cB} \big) \cS^{\cB}{}_{\rho\sigma}
\, \cT^{\rho\sigma}{}_{\cA}\\
&= \cT^{\mu}{}_{\cA} +\big(\hat{\cT}{}^{\mu}{}_{\cB} -
\cT^{\mu}{}_{\cB} \big) \big(\delta^{\cB}_{\cA} -  \cS^{\cB}{}_{\nu}
\, \cT^{\nu}{}_{\cA}\big)\\
&= \hat{\cT}{}^{\mu}{}_{\cA}\ ,
\end{align}
so it is possible to transform $\cT$ such that the new $\tilde{\cT}$ 
coincides in its $\mu$-components with $\hat{\cT}$. This means that it
is always possible to redefine the gauge parameters such that the
projection on pure diffeomorphisms is indeed given by $P$ as defined
in~\eqref{projector}.

Assume now that we have fixed $\cS^{\cA}{}_{\mu}=e^{\cA}_{\mu}$ as well as
$\cT^{\mu}{}_{\cA}$ as in~\eqref{hatT}. Then we are left with block-diagonal transformation
matrices $\Phi$ and $\Psi$ with the identity matrix in the
$\mu-\nu$-block. Suppose now that we have found a matrix
$\hat{\cS}{}^{\cA}{}_{\mu\nu}$ such that
\begin{equation}
\cT^{\rho}{}_{\cA}\, \hat{\cS}{}^{\cA}_{\mu\nu} = 0 \ .
\end{equation}
In our case such a $\hat{\cS}$ is given by the expression
in~\eqref{proposalforSAnurho}. Now we set 
\begin{equation}
\Phi =\begin{pmatrix}
\mathbf{1} & 0\\
0 & *
\end{pmatrix}\quad , \quad 
\Phi^{\mu\nu}{}_{\rho\sigma} = \cT^{\mu\nu}{}_{\cA}\,
\hat{\cS}{}^{\cA}{}_{\rho\sigma} \ .
\end{equation}
With this map $\Phi$, the matrix $\cS$ is transformed to
\begin{align}
\tilde{\cS}{}^{\cA}{}_{\rho\sigma} &= \cS^{\cA}{}_{\mu\nu} \, \Phi^{\mu\nu}{}_{\rho\sigma}\\
&= \cS^{\cA}{}_{\mu\nu} \,\cT^{\mu\nu}{}_{\cB}\,
\hat{\cS}{}^{\cB}{}_{\rho\sigma}\\
&= \big( \delta^{\cA}_{\cB} -  \cS^{\cA}{}_{\mu} \,\cT^{\mu}{}_{\cB}
\big)\hat{\cS}{}^{\cB}{}_{\rho\sigma}\\
&= \hat{\cS}{}^{\cA}{}_{\rho\sigma} \ .
\end{align}
We have to make sure that the transformation $\Phi$ that we defined is
an allowed one, i.e.\ that it does not reduce the set of gauge
transformations. In the case at hand this is clear at least
in a perturbative expansion in the higher-spin fields, where we only
have to check that the transformation is regular at leading
order. The leading terms of $\hat{\cS}$ and $\cS$ coincide and are given by the linearised
expression~\eqref{linearisedgaugetrafo}, therefore the transformation is regular. 

In conclusion we have shown that indeed there is a parameterisation of
the metric-like gauge transformations such that the proposed
map~\eqref{proposedmap} gives an exact relation between metric-like
and frame-like gauge parameters.


\section{Translating frame- to metric-like quantities} 

In this section we will discuss an algorithm to translate frame-like
to metric-like quantities. We will first outline this algorithm for
quantities which do not contain any covariant derivatives and
illustrate it by explicitly calculating the cosmological constant term
in the metric-like formulation up to quartic order in the spin-3 field. We will then generalise the algorithm
appropriately for quantities containing covariant derivatives. This
will allow us to explicitly calculate the gauge transformations of the
metric and spin-3 field to cubic order. Finally we
will discuss why in these cases the mapping between metric-like and
frame-like quantities is unique despite the appearance of seemingly
free parameters in the metric-like expressions.

\subsection{Restricting to vielbeins only}
\label{sec:algorithm}

The aim of this section is to describe an algorithm which allows us to
rewrite a frame-like expression in terms of metric-like fields. This
algorithm is based on a perturbative expansion of all quantities in
the spin-3 field $\phi$. To this end we split the $sl(3)$ generators
into $sl(2)$ generators $\{J_a\}$, labelled by small Latin indices,
and the remaining generators $\{J_A\}$, labelled by capital Latin
indices and chosen to be orthogonal to the $J_a$ with respect to the
Killing form. Using this notation we decompose the vielbein into the
following components
\begin{equation}
e^\mathcal{A} = (e^a,E^A) \ .
\end{equation}
We first note that a given order in the spin-3 field $\phi$ corresponds to the same order of vielbeins $E^A$,
\begin{equation}
\mathcal{O}(\phi) = \mathcal{O}(E) \ .
\end{equation}
This can be seen by expanding (\ref{eq:spin3Field}) and (\ref{eq:metric}),
\begin{align}
\label{eq:phiExpantion}
\phi_{\mu \nu \rho} &= \frac{1}{2} d_{Abc} \,E^A_{(\mu} \,e^b_\nu \,e^c_{\rho)} +
\frac{1}{6} \,d_{ABC} \,E^A_\mu \,E^B_\nu \,E^C_\rho \;, \\ 
\label{eq:metricExpantion}
g_{\mu \nu} &= \kappa_{ab} \,e^a_\mu \,e^b_\nu + \kappa_{AB} \,E^A_\mu
\,E^B_\nu =: \gn_{\mu \nu} +
\overset{\text{\scalebox{.5}{(2)}}}{g}{}_{\mu \nu} \ .
\end{align}
Given any frame-like expression\footnote{For the moment we will not consider terms containing covariant derivatives. But, as discussed in section \ref{sec:algorithmSpinConnection}, by slightly modifying our algorithm these kind of terms can be dealt with as well.}, with space-time indices $\mu_1 \dots \mu_m$ and all frame indices contracted, we can perturbatively find the metric-like equivalent by making an ansatz consisting of all possible contractions of metric-like fields up to a certain order $n$ in the spin-3 field $\phi$. We then proceed in five steps:

\begin{description}
 \item[Step 1:] Expand both sides in terms of $E^A$ using (\ref{eq:metricExpantion}) and (\ref{eq:phiExpantion}) up to order $n$ and subtract them from each other.
 \item[Step 2:] Isolate different orders in $E^A$. For each order we obtain an equation of the following form
\begin{equation}
\label{eq:algorithmStartinPoint}
\sum \limits_i c^{(i)} \,t^{(i)}_{a_1 \dots a_{p_i} B_1 \dots B_{l_i}} (\{e\}, \{ E \}, \{\epsilon \}, \{\gn {}^{-1}\}  )^{a_1 \dots a_{p_i}B_1 \dots B_{l_i}}_{\mu_1 \dots \mu_m} = 0
\end{equation}
where $(\{e\}, \{ E \}, \{\epsilon \}, \{\gn {}^{-1}\}  )$ denotes a contraction of vielbeins of the given index structure containing the inverse zero-order metric, the vielbeins and the invariant space-time tensor $\epsilon^{\mu \nu \rho}$. We will assume that each term has the same number of $\epsilon^{\mu \nu \rho}$, which carries only upper indices. The $t^{(i)}$ are $sl(2)$-invariant tensors. Furthermore some of the $c^{(i)}$ are understood to be the coefficients of the terms arising from the expansion of the frame-like side and are therefore equal to $1$.
The next steps are to be performed for each order separately.

\item[Step 3:] Replace the $E^A$ by
\begin{equation}
\label{eq:resolveEContraction}
E^A_\mu = E^A_\sigma \; \Delta^{\sigma}_{\mu}\ ,
\end{equation}
where we used the following definition
\begin{equation}
\label{eq:capitalDelta}
\Delta^{\sigma}_{\mu}=\gn^{\sigma \rho} \,e^b_\rho \,e^c_\mu \,\kappa_{bc}\;.
\end{equation}
This operation ensures that the spacetime index of $E$ is now contracted with a $sl(2)$-vielbein $e$.

\item[Step 4:] Impose 
\begin{equation}
\label{eq:relKappa}
\gn^{\mu \nu} \,e^a_\mu \,e^b_\nu  = \kappa^{ab}
\end{equation}
for all contractions of this type. After this replacement all terms in the sum of (\ref{eq:algorithmStartinPoint}) are of the same form and can therefore be written as
\begin{equation}
\label{eq:expanSameForm}
\tilde{t}_{a_1 \dots a_{p}B_1 \dots B_{l}}(\{c^{(i)}\}) \; \;  (\{e\}, \{ E \}, \{\epsilon \}, \{\gn {}^{-1}\}  )^{a_1 \dots a_{p}B_1 \dots B_{l}}_{\mu_1 \dots \mu_m} = 0 \ .
\end{equation}
This is because the replacement (\ref{eq:resolveEContraction}) will transfer the space-time index of the vielbein $E$ to a $sl(2)$-vielbein $e$. If the vielbein $E$ carries a free space-time index it is therefore ensured that the free index is now carried by an $sl(2)$-vielbein. If however the space-time indices of two vielbeins $E$ are contracted with each other they will be contracted with a $sl(2)$-vielbein after the substitution (\ref{eq:resolveEContraction}) and imposing (\ref{eq:relKappa}). Finally a vielbein $E$ contracted with an $sl(2)$-vielbein $e$ will stay invariant under both performing (\ref{eq:resolveEContraction}) and (\ref{eq:relKappa}).
Note that the number of $sl(2)$ frame indices in (\ref{eq:expanSameForm}) might have changed during this step. 

\item[Step 5:] Solve (\ref{eq:expanSameForm}) by stripping off the vielbeins. This leads to
\begin{equation}
\label{eq:projLinEq}
\mathcal{P} \tilde{t}_{a_1 \dots a_{p}B_1 \dots B_{l}}(\{ c^{(i)} \}) = 0 \ ,
\end{equation}
where $\mathcal{P}$ is a projector imposing the symmetry inherent in the tensor $(\{e\}, \{ E \}, \{\epsilon \}, \{  \overset{\text{\scalebox{0.5}{(0)}} }{g} {}^{-1}\}  )$. We will explain this aspect in more detail in the next section. But (\ref{eq:projLinEq}) is a linear equation in the coefficients $c^{(i)}$ and can therefore easily be solved using a computer algebra program. 
\end{description}
We stress again that step 3 to 5 have to be performed for all orders from $0$ to $n$ separately.

\subsection{Example: cosmological constant term}

Let us illustrate the algorithm described in the previous section by an example. We will consider the higher-spin cosmological constant term which is given by
\begin{equation}
\label{eq:cosConstTerm}
\frac{1}{3 l^2} \text{tr} \left( e \wedge e \wedge e \right) =
\frac{1}{3 l^2} \,f_{\mathcal{A} \mathcal{B} \mathcal{C}}\,
\epsilon^{\mu \nu \rho}\, e^\mathcal{A}_\mu \,e^\mathcal{B}_\nu
\,e^\mathcal{C}_\rho \; d^3x \ .
\end{equation}
We make an ansatz for the metric-like equivalent of this term by writing down all the possible contractions of the spin-3 field with the metric such that the resulting expression is a space-time scalar, i.e.
\begin{equation}
\label{eq:ccTermAnsatz}
\frac{1}{3 l^2} \,f_{\mathcal{A} \mathcal{B} \mathcal{C}}
\,\epsilon^{\mu \nu \rho} \,e^\mathcal{A}_\mu \,e^\mathcal{B}_\nu
\,e^\mathcal{C}_\rho  = \frac{2}{l^2} \sqrt{-g}  \left( 1 + \sum
\limits_{n=2}^{\infty} \mathcal{L}_n \right) \ .
\end{equation}
Here $\mathcal{L}_n$ denotes all possible contractions compatible with the symmetries of the equation's lhs containing $n$ of the $\phi$ fields and an arbitrary number of metric tensors. In the case of $n=2$ this is given by
\begin{equation}
\mathcal{L}_2 =  c_1 \,\phi^{\mu \nu
\rho} \,\phi_{\mu \nu \rho} + c_2 \,\phi^{\mu} \,\phi_{\mu} \ ,
\end{equation}
where $\phi_\mu$ denotes the trace of the spin-3 field.
We will now explain how the algorithm described in the previous section allows us to fix the coefficients $c_1$ and $c_2$. 

\begin{description}
\item[Step 1 and 2:]We expand (\ref{eq:ccTermAnsatz}) up to second order in $E^A$ which corresponds to second order in $\phi$ as explained in the previous section. For this we have to expand the determinant of the metric which will also depend on $E^A$. This yields
\begin{equation}
\begin{split}
\sqrt{-g} & =  \frac{1}{3!} \,f_{uvw}\, \epsilon^{\chi \delta
\epsilon} \,e^u_\chi \,e^v_\delta \,e^w_\epsilon \left( 1 +
\frac{1}{2} \,\gn^{\mu \nu} \,\kappa_{AB} \,E^A_\mu \,E^B_\nu \right)
+ \mathcal{O}(E^4) \ .
\end{split}
\end{equation}
Subtracting the lhs from the rhs of equation (\ref{eq:ccTermAnsatz}) and considering only terms of quadratic order we obtain up to an overall factor
\begin{equation}
\begin{split}
& c_1 \; f_{u v w} \,d_{Acd} \,d_{Bef} \; \epsilon^{\chi \delta
\epsilon} \,e^u_\chi \,e^v_\delta \,e^w_\epsilon \,e^c_\rho
\,e^e_{\rho'} \,\gn^{\rho \rho'} \,e^d_\sigma \,e^f_{\sigma'}
\,\gn^{\sigma \sigma'} \,E^A_\mu \,E^B_\nu  \,\gn^{\mu \nu} \\
 + & 2 c_1 \; f_{u v w} \,d_{Acd} \,d_{Bef} \; \epsilon^{\chi \delta
 \epsilon} \,e^u_\chi \,e^v_\delta \,e^w_\epsilon \,e^c_\rho
 \,e^e_{\rho'} \,\gn^{\rho \rho'} \,E^A_\sigma \,e^f_{\sigma'}
 \,\gn^{\sigma \sigma'} \,e^d_\mu \,E^B_\nu  \,\gn^{\mu \nu} \\
 + & \ldots \\ + &  \mathcal{O}(E^4) = 0 \ .
\end{split}
\end{equation}
Here we have only written out two terms explicitly and we will now show how the algorithm transforms them into the same form. 

\item[Step 3:]Performing the substitution (\ref{eq:resolveEContraction}) leads to
\begin{equation}
\begin{split}
& c_1 \; f_{u v w}\,  d_{Acd} \,d_{Bef} \; \epsilon^{\chi \delta
\epsilon} \,e^u_\chi \,e^v_\delta \,e^w_\epsilon  \,e^c_\rho
\,e^e_{\rho'} \,\gn^{\rho \rho'} \,e^d_\sigma \,e^f_{\sigma'}
\,\gn^{\sigma \sigma'} \; E^A_\kappa \,\gn^{\kappa \gamma}
\,e^g_\gamma \,e^h_\mu \,\kappa_{g h} \; E^B_\lambda \,\gn^{\lambda
\tau} \,e^l_\tau \,e^j_\nu \,\kappa_{l j} \;  \gn^{\mu \nu} \\ 
+ & 2 c_1 \; f_{u v w}  \,d_{Acd} \,d_{Bef} \; \epsilon^{\chi \delta
\epsilon} \,e^u_\chi \,e^v_\delta \,e^w_\epsilon  \,e^c_\rho
\,e^e_{\rho'} \,\gn^{\rho \rho'} \; E^A_\kappa \,\gn^{\kappa \gamma}
\,e^g_\gamma \,e^h_\sigma \,\kappa_{g h}  \,e^f_{\sigma'}
\,\gn^{\sigma \sigma'} \;e^d_\mu \,E^B_\lambda \,\gn^{\lambda \tau}
\,e^l_\tau \,e^j_\nu \,\kappa_{lj}\,   \gn^{\mu \nu} 
\\ + & \ldots \\ + &  \mathcal{O}(E^4) = 0 \ .
\end{split}
\end{equation}

\item[Step 4:]
Imposing the relation (\ref{eq:relKappa}) we obtain
\begin{equation}
\begin{split}
& c_1 \; f_{u v w}  \,d_{Acd} \,d_{B}{}^{cd} \,\kappa_{lg} \;
\epsilon^{\chi \delta \epsilon} \,e^u_\chi \,e^v_\delta \,e^w_\epsilon
\; E^A_\kappa \,\gn^{\kappa \gamma} \,e^g_\gamma \,E^B_\lambda
\,\gn^{\lambda \tau} \,e^l_\tau 
\\ + & 2 c_1 \; f_{u v w}  \,d_{Adc} \,d_{Bf}{}^{c} \; \; \; \;
\epsilon^{\chi \delta \epsilon} \,e^u_\chi \,e^v_\delta \,e^w_\epsilon
\; E^A_\kappa \,\gn^{\kappa \gamma} \,e^f_\gamma  \,E^B_\lambda
\,\gn^{\lambda \tau} \,e^d_\tau  
\\ + & \ldots \\ + &  \mathcal{O}(E^4) = 0 \ .
\end{split}
\end{equation}
Having written all terms in the same form we find up to a global factor
\begin{equation}
\label{eq:expandedCosConstant}
\begin{split}
&( -108 \,f_{ABc} \,\kappa_{gd} \,\kappa_{he} + f_{cde} (6c_1 \,d_{Ahf}
\,d_{Bg}{}^{f} + 4 c_2 \,d_{Agf} \,d_{Bh}{}^f + 3 c_1
\,d_{Asf} \,d_{B}{}^{sf} \,\kappa_{gh}) 
\\ &+ 18 \,f_{cde} \,\kappa_{AB} \,\kappa_{gh} ) \;  \epsilon^{\chi
\delta \epsilon} \,e^c_\chi \,e^d_\delta \,e^e_\epsilon \; E^A_\gamma
\,\gn^{\alpha \gamma}  \,e^g_\alpha \;  E^B_\sigma  \,\gn^{\beta
\sigma} \,e^h_\beta  = 0 \ .
\end{split}
\end{equation}

\item[Step 5:]
We can solve this equation by stripping off the vielbeins. The remaining term has to be antisymmetrised in $c,d,e$ and symmetrised with respect to exchange of the pair $g,A$ with $h,B$. This operation was denoted by $\mathcal{P}$ in (\ref{eq:projLinEq}). The resulting equation is linear in $c_1$,$c_2$ and can be easily solved, 
\begin{equation}
c_1 = -3 \;, \; c_2 = \frac{9}{2}\ .
\end{equation}
This is most conveniently done by choosing an explicit representation for the invariant tensors of $sl(3)$ and solving the resulting equation using a computer algebra program.
\end{description}
Therefore by using the algorithm described in the previous section we have found the metric-like equivalent of the cosmological constant term to quadratic order in the spin-3 field. 
\\
\\
By applying the algorithm also to the quartic order we obtain the result
\begin{equation}
\label{eq:quadrFinRes}
\frac{1}{3 l^2} \int  f_{\mathcal{A} \mathcal{B} \mathcal{C}} \;
e^\mathcal{A} \wedge e^\mathcal{B} \wedge e^\mathcal{C} =
\frac{2}{l^2} \int d^3x \; \sqrt{-g} \left( 1  + \mathcal{L}_2 +
\mathcal{L}_4 \right) + \mathcal{O}(\phi^6) \ ,
\end{equation}
where the quadratic terms are given by
\begin{equation}
\mathcal{L}_2 =    - 3 \,\phi^{\mu
\nu \rho} \,\phi_{\mu \nu \rho} + \frac{9}{2} \,\phi^{\mu} \,\phi_\mu \ ,
\end{equation}
and the quartic contribution is
\begin{equation}
\label{eq:quadrContrToCosmologicalConstant}
\begin{split}
\mathcal{L}_4 =& (9 + c) \; \phi_{\mu}{}^{\sigma \kappa} \,\phi^{\mu
\nu \rho} \,\phi_{\nu \sigma}{}^{\tau} \,\phi_{\rho \kappa \tau} + c \;
\phi_{\mu \nu}{}^{\sigma} \,\phi^{\mu \nu \rho}
\,\phi_{\rho}{}^{\kappa \tau} \,\phi_{\sigma \kappa \tau} \\ 
& - (54 + 4 c) \;
\phi^{\nu} \,\phi_{\nu}{}^{\rho \sigma} \,\phi_{\rho}{}^{\kappa \tau}
\,\phi_{\sigma \kappa \tau}  - 9 \;
\phi^{\nu} \,\phi_{\nu}{}^{\rho \sigma} \,\phi_{\rho}\, 
\phi_{\sigma} \\
& - (6 +  \tfrac{1}{2} c) \; \phi_{\mu \nu \rho} \,\phi^{\mu \nu\rho}\, 
\phi_{\sigma \kappa \tau} \,\phi^{\sigma \kappa \tau} + (\tfrac{9}{2} + c) \; \phi^{\nu} 
\,\phi_{\nu} \,\phi_{\sigma \kappa \tau} \,\phi^{\sigma \kappa \tau} \\
& + (81 + 2 c) \;
\phi^{\nu} \,\phi_{\nu}{}^{\rho \sigma} \,\phi_{\rho \sigma}{}^{\kappa} 
\,\phi_{\kappa} - ( \tfrac{81}{8} +  \tfrac{1}{2} c) \;
\phi^{\nu} \,\phi_{\nu} \,\phi^{\kappa}\, 
\phi_{\kappa} \ .
\end{split}
\end{equation}
The sum of all terms term proportional to $c$ is zero due to a
dimension dependent identity as will be explained in section~\ref{sec:ambiguities}.

Note that we can not build a scalar by contracting an odd number of spin-3 fields and therefore there are no such contributions in (\ref{eq:quadrFinRes}).

\subsection{Including covariant derivatives}
\label{sec:algorithmSpinConnection}

In the last two sections we did not include terms involving covariant
derivatives in our discussion. In principle we can apply our algorithm
also to these types of terms, but there is an additional
complication. In the frame-like approach covariant derivatives can act
both on $E^A$ and $e^a$.  The algorithm described in section
\ref{sec:algorithm} crucially relies on the fact that we can bring our
expressions into the form (\ref{eq:expanSameForm}).  For this to work
for quantities involving covariant derivatives we need to be able to
express $\mathcal{D}_\mu e^a_\nu$ in terms of $\mathcal{D}_\mu
E^A_\nu$.
This can be achieved as follows. The metric is covariantly constant, 
\begin{equation}
\label{eq:metricCovariantlyConstant}
\nabla_\rho g_{\mu \nu} = \mathcal{D}_\rho g_{\mu \nu} =
\kappa_{\mathcal{A} \mathcal{B}} \,e^\mathcal{A}_\mu
\,\mathcal{D}_\rho e^\mathcal{B}_\nu + \kappa_{\mathcal{A}
\mathcal{B}} \,e^\mathcal{A}_\nu \,\mathcal{D}_\rho e^\mathcal{B}_\mu
= 0 \ ,
\end{equation}
where $\mathcal{D}_\mu$ was defined in (\ref{eq:covariantDer}).

By summing three permutations of equation (\ref{eq:metricCovariantlyConstant}), 
\begin{equation}
\begin{split}
&\kappa_{\mathcal{A} \mathcal{B}} \,e^\mathcal{A}_\mu
\,\mathcal{D}_\rho e^\mathcal{B}_\nu + \kappa_{\mathcal{A}
\mathcal{B}} \,e^\mathcal{A}_\nu \,\mathcal{D}_\rho e^\mathcal{B}_\mu
\\- & \kappa_{\mathcal{A} \mathcal{B}} \,e^\mathcal{A}_\rho
\,\mathcal{D}_\mu e^\mathcal{B}_\nu - \kappa_{\mathcal{A} \mathcal{B}}
\,e^\mathcal{A}_\nu \,\mathcal{D}_\mu e^\mathcal{B}_\rho
\\+& \kappa_{\mathcal{A} \mathcal{B}} \,e^\mathcal{A}_\rho
\,\mathcal{D}_\nu e^\mathcal{B}_\mu + \kappa_{\mathcal{A} \mathcal{B}}
\,e^\mathcal{A}_\mu \,\mathcal{D}_\nu e^\mathcal{B}_\rho = 0 \  ,
\end{split}
\end{equation}
and using torsion constraint (\ref{eq:torsionConstraint}) we conclude
\begin{equation}
\kappa_{\mathcal{A} \mathcal{B}} \; e^\mathcal{A}_\mu \;
\mathcal{D}_\rho e^\mathcal{B}_\nu = 0 \ .
\end{equation}
Expanding this we obtain
\begin{equation}
\mathcal{D}_\mu e^c_\nu = - \kappa_{AB} \,\gn^{\sigma \rho}
\,e^c_\sigma \,E^A_\rho \,\mathcal{D}_\mu E^B_\nu \ .
\end{equation}
Using this result we can reformulate the algorithm described in section \ref{sec:algorithm} such that it is also applicable to expressions involving covariant derivatives.
We only need to modify the prescriptions for step 1 and step 3.

\begin{description}
  \item[Step 1':] Expand both sides in terms of $E^A$ using (\ref{eq:metricExpantion}) and (\ref{eq:phiExpantion}) up to order $n$ and subtract them from each other. Perform the following substitution
  \begin{equation}
	\mathcal{D}_\mu e^c_\nu = - \kappa_{AB} \,\gn^{\sigma \rho}
	\,e^c_\sigma \,E^A_\rho \,\mathcal{D}_\mu E^B_\nu \ .
  \end{equation}
  This ensures that all the covariant derivatives act on $E^A$.
  \item[Step 3':] Replace covariant derivatives of $E^A$ by
  \begin{equation}
  D_\mu E^A_\nu =  \Delta^{\sigma}_{\mu} \, \Delta^{\rho}_{\nu} \, D_\sigma E^A_\rho 
  \end{equation}
  and the $E^A$ without a derivative by
  \begin{equation}
	E^A_\mu = E^A_\sigma \; \Delta^{\sigma}_\mu \ .
  \end{equation}
  The notation $\Delta^{\rho}_{\mu}$ was defined in (\ref{eq:capitalDelta}).
\end{description}
All others steps are unchanged. 

\subsection{Spin-3 transformations}
\label{sec:gaugeTransformations}
In this section we determine the spin-3 transformations of both the metric and the spin-3 field perturbatively. For this we again make the most general ansatz for the gauge transformations of the metric-like fields and fix its coefficients by applying the modified algorithm described in the last section.
The gauge transformation of the spin-3 field is then given by
\begin{equation}
\label{eq:spin3GaugeTrafo}
\delta^{(3)}_\Xi \phi_{\alpha \beta \chi}= 3\,
\nabla_{(\alpha} \hat{\xi}_{\beta \chi)} + (\hat{\xi} \phi \nabla
\phi)_{\alpha \beta \chi} + 
(\nabla \hat{\xi} \phi \phi)_{\alpha \beta\chi} + \mathcal{O}(\phi^4) \ .
\end{equation}
Here $\hat{\xi}$ denotes the traceless component of $\xi$, see (\ref{eq:tracelessProjOfParameter}).
The explicit expressions for $(\hat{\xi} \phi \nabla \phi)_{\alpha \beta \chi}$ and $(\nabla \hat{\xi} \phi \phi)_{\alpha \beta \chi}$ are quite involved and are given in appendix \ref{sec:AppendixGaugeTransformations}.
\\
The metric transforms as follows
\begin{align}
\label{eq:metricGaugeTrafo}
\delta^{(3)}_\Xi g_{\alpha \beta}=&\,6 \,(2\,\hat{\xi}^{\chi\delta}\,\nabla_{\delta} \phi_{\alpha\beta\chi}{}
+4\,\hat{\xi}^{\chi\delta}\,g_{\alpha\beta}\,\nabla_{\delta} \phi_{\chi}{}+\hat{\xi}_{\alpha\beta}\,\nabla_\delta \phi^{\delta}{}
-2\,\hat{\xi}^{\chi\delta}\,g_{\alpha\beta}\,\nabla_\epsilon \phi_{\chi\delta}{}^{\epsilon}{}\nonumber
\\&-4\,\hat{\xi}_{(\alpha}{}^{\chi}\,\triangledown_{\beta)}\phi_{\chi}
-4\,\hat{\xi}_{(\alpha}{}^{\chi}\,\triangledown_{|\chi|}\phi_{\beta)}+
4\,\hat{\xi}_{(\alpha}{}^{\chi}\,\triangledown^{\delta}\phi_{\beta)\chi\delta}\nonumber
\\&-4\,\hat{\xi}^{\chi\delta}\,\triangledown_{(\alpha}\phi_{\beta)\chi\delta}) 
+ (\hat{\xi} \phi \phi \nabla \phi)_{\alpha \beta} +
\mathcal{O}(\phi^5) \ .
\end{align}
The explicit expression for $(\hat{\xi} \phi \phi \nabla \phi)_{\alpha \beta}$ can be found in appendix~\ref{sec:AppendixGaugeTransformations}.

\subsection{Ambiguities}
\label{sec:ambiguities}

Equation (\ref{eq:quadrContrToCosmologicalConstant}) contains a free parameter $c$, seemingly suggesting that the frame-like cosmological constant term does not have a unique metric-like counterpart. However this is not the case. The parameter $c$ is due to a dimensional dependent identity (DDI), which arises by over-antisymmetrisation.  An example for a DDI is
\begin{equation}
\label{eq:masterEqForDDI}
\delta^\mu_{[ \sigma} \,\delta^\nu_{\rho} \,\delta^\gamma_{\sigma}
\,\delta^\lambda_{\tau]} = 0 \ ,
\end{equation}
which obviously vanishes in three dimensions. A systematic way to construct all DDIs of a set of tensors is described in  \cite{Nutma:2013zea}. For a certain tensor all possible contractions with (\ref{eq:masterEqForDDI}) are determined. All identities which arise by over-antisymmetrisation can be constructed in such a way as we can always pull out deltas on the over-antisymmetrised indices. Using the Mathematica package xTras, described in \cite{Nutma:2013zea}, these identities can automatically constructed by this method. For the case of the cosmological constant term at quartic order there is the following relevant DDI,
\begin{equation}
\begin{split}
& \phi_{\mu}{}^{\delta \epsilon} \,\phi^{\mu \nu \chi} \,\phi_{\nu\delta}{}^{\phi} \,\phi_{\chi \epsilon \phi } + 
\phi_{\mu \nu}{}^{\delta}\,\phi^{\mu \nu \chi}\,\phi_{\chi}{}^{\epsilon \phi} \,\phi_{\delta \epsilon \phi} 
- 4 \,\phi^{\nu}\, \phi_{\nu}{}^{\chi \delta}\,\phi_{\chi}{}^{\epsilon \phi}\, \phi_{\delta \epsilon \phi} 
-  \tfrac{1}{2} \,\phi_{\mu \nu \chi} \,\phi^{\mu \nu \chi}\,\phi_{\delta \epsilon \phi} \,\phi^{\delta \epsilon \phi} \\ 
& + \phi^{\nu}\,\phi_{\nu}\,\phi_{\delta \epsilon \phi}\,\phi^{\delta\epsilon \phi}   
+ 2\,\phi^{\nu}\,\phi_{\nu}{}^{\chi \delta}\,\phi_{\chi\delta}{}^{\epsilon}\,\phi_{\epsilon} 
-
\tfrac{1}{2}\,\phi^{\nu}\,\phi_{\nu}\,\phi^{\epsilon}\,\phi_{\epsilon}{}
\equiv 0 \ .
\end{split}
\end{equation}
But the terms proportional to $c$ in (\ref{eq:quadrContrToCosmologicalConstant}) are exactly given by this DDI and therefore vanish. Thus the cosmological constant term to quartic order is uniquely determined by our calculation. 


\section{Gauge algebra} 

We will now discuss the algebra of the gauge transformations for the
metric-like fields. While the algebra of the frame-like
transformations closes off-shell, in the metric-like formulation the
algebra only closes on-shell. We start by explaining this phenomenon
and discuss also why the commutators with spin-2 transformations
(diffeomorphisms) still close off-shell. We then determine explicitly
the gauge algebra to linear order in the spin-3 field $\phi$.

\subsection{On-shell gauge algebra}

Recall from section \ref{sec:proposalForTheMap} that in the frame-like
theory general local translations induce pure diffeomorphisms and
spin-3 transformations. General translations are parametrised by
$\Xi^\mathcal{A}$. Obviously this corresponds to having 8 degrees of
freedom which nicely matches the 3+5 degrees of freedom of the
parameter $v^\mu$ corresponding to pure diffeomorphisms and
$\hat{\xi}^{\mu \nu} = \xi^{\mu \nu} - \frac{1}{3} \,g^{\mu \nu}\,
\xi^{\lambda}{}_{\lambda} $ parameterising the spin-3
transformations. According to (\ref{eq:genLocalTranslations}) the
fields of the frame-like formalism transform as follows
\begin{align}
\label{eq:frameLikeGaugeTrafo}
\delta_{\Xi} e_{\mu}^{\mathcal{A}} & = D_{\mu} \Xi^{\mathcal{A}} \ , \\
\delta_{\Xi} \omega_{\mu}^{\mathcal{A}} & = \frac{1}{l^2} \left[
e_\mu, \Xi^{\mathcal{A}} \right] \ ,
\label{spinconnectionGaugeTrafo}
\end{align}
and the gauge algebra closes off-shell.

When we translate the frame-like theory to the metric-like formulation
we have to use the torsion constraint
(\ref{eq:torsionConstraint}) to express the spin connection in terms
of vielbeins, $\omega=\omega(e)$. This implicit dependence induces a
gauge transformation of the spin connection that differs from the
transformation~\eqref{spinconnectionGaugeTrafo}, and only coincides
with it on-shell, i.e.\ after using the equation of motion.
This can be seen as follows. The induced transformation of the spin
connection can be calculated by varying the torsion
constraint~\eqref{eq:torsionConstraint},
\begin{equation}
\delta_\Xi \left( D_{[\mu } e_{\nu ]}^\mathcal{A} \right) = \delta_\Xi
D_{[\mu} e_{\nu ]}^\mathcal{A} +D_{[\mu } D_{\nu ]} \Xi^\mathcal{A} =
0 \ .
\end{equation}
The Christoffel symbol is symmetric in $\mu$ and $\nu$ and therefore the
variation of the covariant derivative in the equation above is given by the transformation
of the spin connection. We thus obtain
\begin{equation}
\label{eq:variationOfSpinConnection}
f^{\mathcal{A}}{}_{\mathcal{B}\mathcal{C}} \,\delta_\Xi
\omega^\mathcal{B}_{[\mu} \,e_{\nu ]}^\mathcal{C} +
f^{\mathcal{A}}{}_{\mathcal{B}\mathcal{C}}\, R_{\mu \nu}^\mathcal{B}
\,\Xi^\mathcal{C} = 0
\end{equation}
with $R^\mathcal{A}_{\mu \nu}$ from (\ref{eq:curvatureTensor}). The
equation of motion for the vielbein is 
\begin{equation}\label{eqofmotion}
R^\mathcal{A}_{\mu \nu}=-\frac{1}{2l^2}
f^\mathcal{A}{}_{\mathcal{B}\mathcal{C}}\,e^\mathcal{B}_\mu\,
e^\mathcal{C}_\nu \ .
\end{equation}
Using this equation of motion in~\eqref{eq:variationOfSpinConnection},
we find that the induced transformation reduces on-shell to (we assume that the vielbein is
non-degenerate) 
\begin{equation}
\delta_\Xi \omega^\mathcal{A}_{\mu} = \frac{1}{l^2}
f^\mathcal{A}{}_{\mathcal{B}\mathcal{C}}\,e^\mathcal{B}_\mu\,
\Xi^\mathcal{C} \ ,
\end{equation}
which coincides with the
transformation~\eqref{spinconnectionGaugeTrafo} in the frame-like
theory. Therefore we expect that the metric-like gauge algebra only
closes on-shell.

Let us explicitly consider the commutator of two gauge transformations
on a vielbein (all metric-like fields are built out of the vielbein).
Using (\ref{eq:frameLikeGaugeTrafo}) we obtain
\begin{align}
[\delta_\Xi, \delta_\Pi] e^\mathcal{A}_\mu &= D_\mu \left( \delta_\Xi \Pi^\mathcal{A} - \delta_\Pi \Xi^\mathcal{A} \right) + f^{\mathcal{A}}{}_{\mathcal{B}\mathcal{C}} \left( \delta_\Xi \omega^\mathcal{B}_{\mu} \,\Pi^\mathcal{C} - \delta_\Pi \omega^\mathcal{B}_{\mu} \,\Xi^\mathcal{C} \right) \nonumber \\
&= \delta_{(\delta_\Xi\Pi-\delta_\Pi\Xi)} e_{\mu}^{\cA} +
f^{\mathcal{A}}{}_{\mathcal{B}\mathcal{C}} \left( \delta_\Xi
\omega^\mathcal{B}_{\mu} \,\Pi^\mathcal{C} - \delta_\Pi
\omega^\mathcal{B}_{\mu} \,\Xi^\mathcal{C} \right) \ .\label{eq:vielbeinCommuator}
\end{align}
The first term is a local translation of the vielbein and therefore
can again be interpreted as a gauge transformation in the metric-like
formulation. For the second term it might in general not be possible
to rewrite it as a gauge transformation on the vielbein. On the other
hand, on-shell the last term is a generalised local Lorentz transformation of the
vielbein as can be checked by using the Jacobi identity,
\begin{equation}
\begin{split}
f^{\mathcal{A}}{}_{\mathcal{B}\mathcal{C}} \left( \delta_\Xi
\omega^\mathcal{B}_{\mu} \,\Pi^\mathcal{C} - \delta_\Pi
\omega^\mathcal{B}_{\mu} \,\Xi^\mathcal{C} \right) &= - \frac{1}{l^2}
f^{\mathcal{A}}{}_{\mathcal{B}\mathcal{C}} \left(
f^\mathcal{B}{}_{\mathcal{D}\mathcal{E}} \,e^\mathcal{D}_\mu\,
\,\Xi^\mathcal{E} \,\Pi^\mathcal{C} -
f^\mathcal{B}{}_{\mathcal{D}\mathcal{E}} \,e^\mathcal{D}_\mu\,
\Pi^\mathcal{E} \,\Xi^\mathcal{C} \right) \\ &= \frac{1}{l^2}
f^\mathcal{A}{}_{\mathcal{B}\mathcal{D}} \left(
f^\mathcal{B}{}_{\mathcal{E}\mathcal{C}}\,\Xi^\mathcal{E}\,
\Pi^\mathcal{C} \right)  e^\mathcal{D}_\mu \ . 
\end{split}
\end{equation}
In the metric-like fields all frame indices are contracted with
invariant tensors, and the local Lorentz transformations do not
have any effect. Hence we find that on-shell the gauge algebra in the
metric-like formulation is obtained by translating
\begin{equation}
\label{eq:commutator}
[\delta_\Xi, \delta_\Pi] =  \delta_{(\delta_\Xi\Pi-\delta_\Pi\Xi)}
\end{equation}
into metric-like quantities.

\subsection{Off-shell closure for spin-2 transformations}
\label{sec:offshellClosureSpin2Transformation}

In the last subsection we have shown that after imposing the
vielbein's equation of motion the second term in
(\ref{eq:vielbeinCommuator}) can be written as a local Lorentz
transformation. In this section we will show that in the special case
in which at least one of the parameters describes a spin-2
transformation, i.e. $\Xi^\mathcal{A}=e^\mathcal{A}_\mu \xi^\mu$, the
last term of (\ref{eq:vielbeinCommuator}) is a local Lorentz
transformation even off-shell.  Firstly, using
(\ref{eq:variationOfSpinConnection}) we calculate the variation of the
spin connection in this special case.
\begin{equation}
\label{eq:tmpSpin2spinConnectionVar}
\begin{split}
f^{\mathcal{A}}{}_{\mathcal{B}\mathcal{C}} \,\delta_\Xi
\omega^\mathcal{B}_{[\mu} \,e_{\nu ]}^\mathcal{C}
&=-f^{\mathcal{A}}{}_{\mathcal{B}\mathcal{C}} \, 
R_{\mu\nu}^\mathcal{B} \, e^\mathcal{C}_\sigma \,\xi^\sigma\\
& =  -f^{\mathcal{A}}{}_{\mathcal{B}\mathcal{C}}  \left( R_{\mu
\sigma}^\mathcal{B} \,e^\mathcal{C}_\nu + R_{\sigma \nu}^\mathcal{B}
\,e^\mathcal{C}_\mu \right) \xi^\sigma \\
& =  2 \,f^{\mathcal{A}}{}_{\mathcal{B}\mathcal{C}}  \, R_{\sigma [
\mu}^\mathcal{B} \,e^\mathcal{C}_{\nu]} \,\xi^\sigma \ ,
\end{split}
\end{equation}
where we have used a Bianchi-like identity
$f^{\mathcal{A}}{}_{\mathcal{B}\mathcal{C}} \,R^\mathcal{B}_{[\mu \nu}
\,e^\mathcal{C}_{\sigma]} = 0$ (see appendix~\ref{app:bianchi}).
As the vielbein is non-degenerate we conclude from (\ref{eq:tmpSpin2spinConnectionVar}) that
\begin{equation}
\delta_\Xi \omega^\mathcal{B}_\mu = 2 \,\xi^\sigma \,R^\mathcal{B}_{\sigma \mu}
\end{equation}
is the induced transformation of the spin connection under a spin-2 transformation.\\
Plugging this result into the second term of
(\ref{eq:vielbeinCommuator}) and using
(\ref{eq:variationOfSpinConnection}) we obtain 
\begin{equation}
\begin{split}
f^{\mathcal{A}}{}_{\mathcal{B}\mathcal{C}} \left( \delta_\Xi
\omega^\mathcal{B}_{\mu} \,\Pi^\mathcal{C} - \delta_\Pi
\omega^\mathcal{B}_{\mu} \,\Xi^\mathcal{C} \right)
&=f^{\mathcal{A}}{}_{\mathcal{B}\mathcal{C}} \left( 2\, \xi^\nu
\,R^\mathcal{B}_{\nu \mu} \,\Pi^\mathcal{C} - \delta_\Pi
\omega_\mu^\mathcal{B} \,e^\mathcal{C}_\nu \,\xi^\nu \right) \\
&= f^{\mathcal{A}}{}_{\mathcal{B}\mathcal{C}} \left( 2\, \xi^\nu
\,R^\mathcal{B}_{\nu \mu} \,\Pi^\mathcal{C} - 2 \,\delta_\Pi
\omega^\mathcal{B}_{[\mu} \,e^\mathcal{C}_{\nu]} \,\xi^\nu  -
\delta_\Pi \omega^\mathcal{B}_{\nu} \,e^\mathcal{C}_{\mu} \,\xi^\nu \right) \\
& = f^{\mathcal{A}}{}_{\mathcal{B}\mathcal{C}} \left( 2\, \xi^\nu
 \,R^\mathcal{B}_{\nu \mu} \,\Pi^\mathcal{C} + 2\,  \xi^\nu
 \,R^\mathcal{B}_{\mu \nu} \,\Pi^\mathcal{C} - \delta_\Pi
 \omega^\mathcal{B}_{\nu} \,e^\mathcal{C}_{\mu} \,\xi^\nu \right) \\
&=  - f^{\mathcal{A}}{}_{\mathcal{B}\mathcal{C}} \,\delta_\Pi
\omega^\mathcal{B}_{\nu} \,e^\mathcal{C}_{\mu} \,\xi^\nu \ .
\end{split}
\end{equation}
But the final expression is just a generalised local Lorentz
transformation and we have therefore shown that the commutator of a
spin-2 transformation with any other transformation can be expressed
as a gauge transformation also off-shell. 

In the following we will compute the various commutators that arise in
the algebra of metric-like gauge transformations explicitly.

\subsection{Spin-2 spin-2 commutator} 

Here we will consider the case of both transformations being
diffeomorphisms, i.e. $\Pi^\mathcal{A} = e^\mathcal{A}_\mu \pi^\mu$ 
and $\Xi^\mathcal{A}=e^\mathcal{A}_\mu \xi^\mu$. As shown in the
previous section this commutator closes off-shell and using
(\ref{eq:vielbeinCommuator}) we can calculate the resulting
transformation
\begin{equation}
\begin{split}
\delta_\Pi \left( e^\mathcal{A}_\mu \,\xi^\mu \right) - \delta_\Xi
\left( e^\mathcal{A}_\mu \,\pi^\mu \right) & =  D_\mu \left(
e^\mathcal{A}_\nu \,\pi^\nu \right) \xi^\mu - \xi \leftrightarrow \pi
\\ &= -e^\mathcal{A}_\nu \,\mathcal{L}_\pi \xi^\nu  + 2 \,\xi^\mu
\,\pi^\nu \,D_{[\mu} e^\mathcal{A}_{\nu]} 
\end{split}
\end{equation} 
where $\mathcal{L}_\pi \xi^\nu = \pi^\mu \partial_\mu \xi^\nu -
\xi^\mu \partial_\mu \pi^\nu$ is the Lie derivative. But the last
term in the last line vanishes as we impose the torsion constraint
(\ref{eq:torsionConstraint}). By (\ref{purediffeo}) the result of this commutator therefore induces a diffeomorphism with vector field $-\mathcal{L}_\pi \xi^\nu$.  

\subsection{Spin-3 spin-2 commutator}

We now want to discuss the commutator of a spin-3 and a spin-2 transformation. The spin-3 transformation is parameterised by
\begin{equation}
\Xi^{\cA} = \cS^{\cA}{}_{\mu_{1}\mu_{2}}\, \xi^{\mu_{1}\mu_{2}} \ ,
\end{equation}
where $\cS$ is given in~\eqref{proposalforSAnurho}. The result for the commutator will not depend on the precise form of $\cS$, but only on the property that it is built from the vielbeins. In fact we can also consider the more general case of the commutator of a spin-$(s+1)$ and a spin-2 transformation without any additional complication, where the spin-2 and the spin-$(s+1)$ transformations are parameterised by
\begin{equation}\label{spin2spins}
\Pi^\mathcal{A} = e^\cA_\sigma \pi^\sigma \; \; \; \text{and}  \; \;
\; \Xi^\mathcal{A} = \cS^\cA{}_{\mu_1 \dots \mu_s} (e) \xi^{\mu_1 \dots \mu_s}
\ .
\end{equation}
Here, $\cS^\mathcal{A}{}_{\mu_1 \dots \mu_s}(e)$ is built by contracting vielbeins and it is completely symmetric in all space-time indices.
For a result that we need later we consider the following space-time tensor,
\begin{equation}
\mathcal{O}_{\nu \mu_1 \dots \mu_s} = \kappa_{\mathcal{A} \mathcal{B}}
\;e^\mathcal{A}_\nu \; \cS^\mathcal{B}{}_{\mu_1 \dots \mu_s}(e) \ .
\end{equation} 
Because it is constructed from the vielbeins, under the spin-2 transformation the tensor $\mathcal{O}_{\nu \mu_1 \dots \mu_s}$ changes by the Lie derivative along $\pi$,
\begin{equation}
\label{eq:transTArhs}
\delta_\Pi \mathcal{O}_{\nu \mu_1 \dots \mu_s}  = \pi^\sigma
\nabla_\sigma \mathcal{O}_{\nu \mu_1 \dots \mu_s} + s \;
\nabla_{(\mu_1} \pi^\sigma \mathcal{O}_{|\nu \sigma| \mu_2 \dots\mu_s)} 
+ \nabla_\nu \pi^\sigma \; \mathcal{O}_{\sigma \mu_1 \dots\mu_s}  \ .
\end{equation}
The lhs of this equation can be calculated by explicitly evaluating the variation of the vielbein, i.e.
\begin{equation}
\label{eq:transTAlhs}
\begin{split}
\delta_\Pi \mathcal{O}_{\nu \mu_1 \dots \mu_s}  
&= \kappa_{\mathcal{A}\mathcal{B}} \,D_\nu (e^\mathcal{A}_\sigma \pi^\sigma)
\cS^\mathcal{B}{}_{\mu_1 \dots \mu_s} + \kappa_{\mathcal{A} \mathcal{B}}\,
e^\mathcal{A}_\nu (\delta_\Pi \cS^\mathcal{B}{}_{\mu_1 \dots \mu_s}) \\ &=
\kappa_{\mathcal{A} \mathcal{B}} \,\pi^\sigma (D_\sigma
e^\mathcal{A}_\nu ) \cS^\mathcal{B}{}_{\mu_1 \dots \mu_s} +
\kappa_{\mathcal{A} \mathcal{B}} \,e^\mathcal{A}_\nu (\delta_\Pi
\cS^\mathcal{B}{}_{\mu_1 \dots \mu_s}) + \nabla_\nu \pi^\sigma
\mathcal{O}_{\sigma \mu_1 \dots \mu_s} \ ,
\end{split}
\end{equation}
where we used (\ref{eq:torsionConstraint}) and suppressed the dependency of $\cS^\mathcal{B}{}_{\mu_1 \dots \mu_s}$ on the vielbeins to simplify notation.
Combining (\ref{eq:transTArhs}) with (\ref{eq:transTAlhs}) yields
\begin{equation}
\kappa_{\mathcal{A} \mathcal{B}} \,e^\mathcal{A}_\nu (\delta_\Pi
\cS^\mathcal{B}{}_{\mu_1 \dots \mu_s}) 
= \kappa_{\mathcal{A}\mathcal{B}}\,
e^\mathcal{A}_\nu \, \pi^\sigma D_\sigma 
\cS^\mathcal{B}{}_{\mu_1 \dots\mu_s} 
+ s \; \kappa_{\mathcal{A}\mathcal{B}} \,e^\mathcal{A}_\nu \;
\nabla_{(\mu_1} \pi^\sigma  \; 
\cS^\mathcal{B}{}_{|\sigma| \mu_2 \dots\mu_s)} \ .
\end{equation}
We therefore conclude that
\begin{equation}
\delta_\Pi \cS^\mathcal{B}{}_{\mu_1 \dots \mu_s} = \pi^\sigma D_\sigma
\cS^\mathcal{B}{}_{\mu_1 \dots \mu_s} + s \; \nabla_{(\mu_1} \pi^\sigma
\; \cS^\mathcal{B}{}_{|\sigma| \mu_2 \dots \mu_s)} \ .
\end{equation}
We are now in the position to determine the commutator of the spin-2 transformation $\Pi$ and the spin-$(s+1)$ transformation $\Xi$ given in~\eqref{spin2spins}, and we find 
\begin{equation}
\begin{split}
\delta_\Xi \Pi^\mathcal{A} - \delta_\Pi \Xi^\mathcal{A} & = \pi^\sigma D_\sigma \left( \cS^\cA{}_{\mu_1 \dots \mu_s} \xi^{\mu_1 \dots \mu_s} \right) -  \xi^{\mu_1 \dots \mu_s} \; \delta_\Pi \cS^\cA{}_{\mu_1 \dots \mu_s} \\ 
&= \cS^\cA{}_{\mu_1 \dots \mu_s} \left( \pi^\sigma \nabla_\sigma \xi^{\mu_1 \dots \mu_s} - s \; \xi^{\sigma (\mu_1 \dots \mu_{s-1}} \nabla_\sigma \pi^{\mu_s)} \; \right) \\ 
&= \cS^\cA{}_{\mu_1 \dots \mu_s} \left( \mathcal{L}_\pi \xi^{\mu_1 \dots\mu_s} \right) \ .
\end{split}
\end{equation}
Thus the commutator is a spin-$(s+1)$ transformation whose parameter is given by the Lie derivative of the original spin-$(s+1)$-parameter. In particular in our case we find
\begin{equation}
[\delta^{(3)}_{\xi},\delta^{(2)}_{\pi}] = \delta^{(3)}_{\mathcal{L}_{\pi}\xi} \ .
\end{equation}

\subsection{Spin-3 spin-3 commutator} 

In contrast to the commutation relation involving at least one spin-2 transformation we currently do not have an all
order result for the commutator of two spin-3 transformations. The commutator is specified by traceless parameters $\hat{\xi}^{\mu \nu}$ and $\hat{\pi}^{\mu \nu}$, and generically it will lead to a combination of a spin-2 transformation and a spin-3 transformation, i.e.
\begin{equation}
\left[ \delta_\Pi, \delta_\Xi \right] e^\mathcal{A}_\mu =
\delta_{\cS(u,v)} \, e^\mathcal{A}_\mu \;,
\end{equation}
where 
\begin{equation}
\label{eq:map2}
\cS^\cA(u,v)=\cS^\cA{}_\mu v^\mu + \cS^\cA{}_{\rho \sigma} u^{\rho \sigma}
\end{equation}
denotes the map defined in (\ref{eq:map}). 
In the following we will determine the parameters $u^{\mu \nu}$ and $v^\mu$ perturbatively in the spin-3 field. First we will calculate explicitly the spin-2 parameter $v^\mu$ by only considering zeroth order contributions. Then we will use the algorithm discussed in section \ref{sec:algorithm} to determine these parameters at linear order.

\subsubsection{Spin-2 parameter $v^\mu$}
\label{sec:spin2}

This contribution was already calculated in \cite{Campoleoni:2012hp} using a different method. We need to evaluate
\begin{equation}
\label{eq:diffStartSpin3Spin3}
\begin{split}
\delta^{(3)}_\Pi \Xi^\mathcal{A}  = & 3 \;
\delta^{(3)}_\Pi \left\{ \delta^\mathcal{A}_\mathcal{B}
- P^\mathcal{A}_\mathcal{B} \right \} \; 
d^\mathcal{B}{}_{\mathcal{C}\mathcal{D}} \;  e^\mathcal{C}_\mu
\,e^\mathcal{D}_\nu \,\hat{\xi}^{\mu \nu}  
+ 3 \left( \delta^\mathcal{A}_\mathcal{B} - P^\mathcal{A}_\mathcal{B}
\right) \; 
\delta^{(3)}_\Pi \left \{ 
d^\mathcal{B}{}_{\mathcal{C}\mathcal{D}} \,e^\mathcal{C}_\mu
\,e^\mathcal{D}_\nu \,\hat{\xi}^{\mu \nu} \right \} 
\\  = & - 3 \left( \delta^{(3)}_\Pi
P^\mathcal{A}_\mathcal{B} \right) \; 
d^\mathcal{B}{}_{\mathcal{C}\mathcal{D}} \;  e^\mathcal{C}_\mu \,e^\mathcal{D}_\nu \,\hat{\xi}^{\mu \nu} 
\\ & + 6 \left( \delta^\mathcal{A}_\mathcal{B} -
P^\mathcal{A}_\mathcal{B} \right) \;  d^\mathcal{B}{}_{\mathcal{C}
\mathcal{D}} \left(  D_\mu \Pi^\mathcal{C} \right) e^\mathcal{D}_\nu
\,\hat{\xi}^{\mu \nu} 
\\& + 3 \left( \delta^\mathcal{A}_\mathcal{B} -
P^\mathcal{A}_\mathcal{B} \right)  \; 
d^\mathcal{B}{}_{\mathcal{C}\mathcal{D}} \; e^\mathcal{C}_\mu \,e^\mathcal{D}_\nu \left(
\delta^{(3)}_\Pi  \hat{\xi}^{\mu \nu} \right)  \ .
\end{split}
\end{equation}
The variation of the projector $P^\mathcal{A}_\mathcal{B}$ is given by
\begin{equation}
\begin{split}
\delta^{(3)}_\Pi P^\mathcal{A}_\mathcal{B} = & \left(
D_\mu \Pi^\mathcal{A} \right) g^{\mu \nu} \,e^\mathcal{C}_\nu \;
\kappa_{\mathcal{B} \mathcal{C}} + e^\mathcal{A}_\mu \,g^{\mu \nu}
\kappa_{\mathcal{B} \mathcal{C}} \left( D_\nu \Pi^\mathcal{C} \right)
\\ & -2 e^\mathcal{A}_\mu \; \kappa_{\mathcal{E} \mathcal{F}} \;
e^\mathcal{C}_\nu\, e^\mathcal{E}_{( \sigma} \,D_{\rho )}
\Pi^\mathcal{F}\,
g^{\sigma \nu} \,g^{\rho \mu}  \; \kappa_{\mathcal{B} \mathcal{C}} \ ,
\end{split}
\end{equation}
where the last term arises due to the variation of the inverse metric in the projector. We will now evaluate (\ref{eq:diffStartSpin3Spin3}) at leading order. Let us focus on the last term in (\ref{eq:diffStartSpin3Spin3}) first. By using (\ref{eq:metricGaugeTrafo}) it can be checked easily that this term is of higher order as
\begin{equation}
\delta^{(3)}_\Pi  \hat{\xi}^{\mu \nu} = \mathcal{O}(E) \ .
\end{equation}
Note that if we choose $\mathcal{A}=A$ all terms in (\ref{eq:diffStartSpin3Spin3}) will be at least of linear order. For $\mathcal{A}=a$ we can easily deduce that the second term in (\ref{eq:diffStartSpin3Spin3}) does not contribute as
\begin{equation}
(\delta^a_b - P^a_b) = 0 \; \text{and} \; P^a_B=\mathcal{O}(E) \ .
\end{equation}
So only the first term in (\ref{eq:diffStartSpin3Spin3}) will contribute to leading order. From $\mathcal{A}=a$ it follows that to leading order we have to choose $\mathcal{B}=B$. The variation of the projector is then given by
\begin{equation}
\delta^{(3)}_\Pi P^a_B  = 3 \,e^a_\mu \,g^{\mu \nu} \;
\kappa_{BC} \,d^{C}{}_{ef} \; e^e_\rho \,e^f_\tau \,\nabla_\nu
\hat{\pi}^{\rho \tau} + \mathcal{O}(E) \ .
\end{equation}
Plugging this in the only non-vanishing term of (\ref{eq:diffStartSpin3Spin3}) we find at leading order
\begin{equation}
\label{eq:spin2spin3CommutatorFinalResult}
\begin{split}
\delta^{(3)}_\Pi \Xi^a -  \delta^{(3)}_\Xi
\Pi^a & = -9 \; d_{Bef} \,d^{B}{}_{cd} \; e^a_\sigma\,
g^{\sigma\tau}\,e^e_\rho\,e^f_\gamma\,e^c_\mu\,e^d_\nu \left(
\hat{\xi}^{\mu \nu} \,\nabla_\tau \hat{\pi}^{\rho \gamma} -  \hat{\xi} \leftrightarrow \hat{\pi}  \right) 
\\ & = - 18 \,e^a_\sigma \,g^{\sigma \tau} \left( \hat{\xi}^{\mu \nu}\,
\nabla_\tau \hat{\pi}_{\mu \nu}  - \hat{\xi} \leftrightarrow \hat{\pi}
\right)  \ , 
\end{split}
\end{equation}
where we have used the identity (\ref{eq:dd-identity}) in the last step.

But by (\ref{purediffeo}) the result in (\ref{eq:spin2spin3CommutatorFinalResult}) corresponds to a spin-2 transformation with the parameter
\begin{equation}
\label{eq:spin2Parameter}
v^\mu= -18 \,g^{\mu \nu} \left( \xi^{\rho \sigma} \,\nabla_\nu 
\pi_{\rho\sigma} - \frac{1}{3} \,\xi^{\rho}{}_{\rho} \,\nabla_\nu
\pi^{\sigma}{}_{\sigma}  -  \xi \leftrightarrow \pi \right) \ .
\end{equation}

\subsubsection{Spin-3 parameter $u^{\mu \nu}$}
\label{sec:spin3}
To determine the spin-3 parameter $u^{\mu \nu}$ we make the following ansatz
\begin{equation}
\label{eq:ansatzSpin3Spin3Spin3}
\delta^{(3)}_\Pi \Xi^\mathcal{A} - \delta^{(3)}_\Xi \Pi^\mathcal{A} = \cS^\mathcal{A}(u^{\mu \nu},v^\mu) \ ,
\end{equation}
where $\cS^\cA$ is defined as in (\ref{eq:map2}). The parameter $v^\mu$ cannot be corrected by terms linear in the spin-3 field as we cannot build a vector by contracting a spin-3 field, a covariant derivative and the parameter $u^{\mu \nu}$. In order to solve this equation we make an ansatz for the linear order of $u^{\mu \nu}$ by considering all possible contractions of 
\begin{equation}
\xi^{\rho \sigma} \; , \; \pi^{\rho \sigma} \; \text{and} \; \phi_{\sigma \tau \rho}
\end{equation}
with two symmetric free indices, $\mu$ and $\nu$, and antisymmetric with respect to the exchange of $\xi$ and $\pi$. We use the algorithm described in section \ref{sec:algorithmSpinConnection} to determine the coefficients of the ansatz. 
The result for $u^{\alpha \beta}$ contains three different contributions denoted by
\begin{equation}
\label{eq:gaugeAlgebraParameter}
u^{\alpha \beta} = u^{\alpha \beta}_1 + u^{\alpha \beta}_2 + u^{\alpha\beta}_3 \ .
\end{equation}
 Firstly terms with a derivative acting on the spin-3 field
\begin{align}
u_1^{\alpha \beta}=&6\bigl(-5(\hat{\xi}\hat{\pi})^{\alpha\beta\chi\delta}\triangledown_{\delta}\phi_{\chi}+(\hat{\xi}\hat{\pi})^{\chi\delta}{}_{\chi}{}^{\epsilon}(g^{\alpha\beta}\triangledown_{\epsilon}\phi_{\delta}-\triangledown_{\epsilon}\phi^{\alpha\beta}{}_{\delta})+3(\hat{\xi}\hat{\pi})^{\alpha\beta\chi\delta}\triangledown_{\epsilon}\phi_{\chi\delta}{}^{\epsilon} \nonumber\\&
-6(\hat{\xi}\hat{\pi})^{\chi\delta\epsilon\xi}g^{\alpha\beta}\triangledown_{\xi}\phi_{\chi\delta\epsilon}+3(\hat{\xi}\hat{\pi})^{(\alpha|\chi}{}_{\chi}{}^{\delta|}\triangledown^{\beta)}\phi_{\delta}+2(\hat{\xi}\hat{\pi})^{(\alpha|\chi}{}_{\chi}{}^{\delta|}\triangledown_{\delta}\phi^{\beta)} \nonumber\\&
-3(\hat{\xi}\hat{\pi})^{(\alpha|\chi}{}_{\chi}{}^{\delta}\triangledown^{\epsilon|}\phi^{\beta)}{}_{\delta\epsilon}-5(\hat{\xi}\hat{\pi})^{(\alpha|\chi\delta\epsilon|}\triangledown_{\chi}\phi^{\beta)}{}_{\delta\epsilon}+13(\hat{\xi}\hat{\pi})^{(\alpha|\chi\delta\epsilon|}\triangledown_{\delta}\phi^{\beta)}{}_{\chi\epsilon}\bigr)
+ k_1 D_1^{\alpha \beta} \ ,
\end{align}
where we have used the following notation
\begin{equation}
\label{eq:commutatorXiPiParameter}
(\hat{\xi} \hat{\pi} )^{\mu \nu \rho \sigma}=
\hat{\xi}^{\mu\nu}\hat{\pi}^{\rho \sigma} - \hat{\xi} \leftrightarrow \hat{\pi} \ .
\end{equation}
The hatted tensors again denote the traceless components of the parameters, see (\ref{eq:tracelessProjOfParameter}). The term $D_1^{\alpha \beta}$ is given in appendix \ref{sec:AppendixDDIContrToGaugeAlgebra} and vanishes due to DDIs.
\\
Secondly there are contributions with a derivative acting on one of the parameters,
\begin{equation}
\begin{split}
u^{ \alpha\beta }_2=&12\left( (\nabla\hat{\xi}\hat{\pi})^{\delta\alpha\beta}{}_{\chi\delta}\phi^{\chi}+(\nabla\hat{\xi}\hat{\pi})^{\delta}{}_{\chi}{}^{\epsilon}{}_{\delta\epsilon}\phi^{\chi}g^{\alpha\beta}-(\nabla\hat{\xi}\hat{\pi})^{\chi\delta\epsilon}{}_{\chi\delta}\phi^{\alpha\beta}{}_{\epsilon}-\tfrac{3}{2}(\nabla\hat{\xi}\hat{\pi})^{\chi\alpha\beta\delta\epsilon}\phi_{\chi\delta\epsilon} \right. \\&
+12(\nabla\hat{\xi}\hat{\pi})^{\chi\delta\epsilon}{}_{\chi}{}^{\xi}g^{\alpha\beta}\phi_{\delta\epsilon\xi}-(\nabla\hat{\xi}\hat{\pi})^{\chi(\alpha|\delta|\beta)}{}_{\chi}\phi_{\delta}-(\nabla\hat{\xi}\hat{\pi})^{\chi(\alpha|\delta|}{}_{\chi\delta}\phi^{\beta)} \\&
\left. -2(\nabla\hat{\xi}\hat{\pi})^{\chi(\alpha|\delta}{}_{\chi}{}^{\epsilon|}\phi^{\beta)}{}_{\delta\epsilon}-34(\nabla\hat{\xi}\hat{\pi})^{\chi\delta\epsilon(\alpha}{}_{\chi}\phi^{\beta)}{}_{\delta\epsilon}
\right) + k_2 D_2^{\alpha \beta} + k_3 D_3^{\alpha \beta} + k_4
D_4^{\alpha \beta} + k_5 D_5^{\alpha \beta} \ .
\end{split}
\end{equation}
Here we used the notation
\begin{equation}
\label{eq:commutatorNablaXiPiParameter}
( \nabla \hat{\xi}  \hat{\pi} )_{\mu}{}^{\nu \rho \sigma \epsilon}=
\hat{\pi}^{\nu \rho} \nabla_\mu \hat{\xi}^{\sigma \epsilon} -
\hat{\pi} \leftrightarrow \hat{\xi} \ .
\end{equation}
The terms $D_i^{\alpha \beta}$, $i=2\dots5$, are given in appendix \ref{sec:AppendixDDIContrToGaugeAlgebra} and are identically zero due to DDIs.
\\
Finally there are contributions containing the trace of the parameters of the gauge transformations
\begin{equation}
\begin{split}
u_3^{\alpha \beta} =& 4(\hat{\xi}\pi'-\hat{\pi}\xi')^{\lambda\rho}\phi^{\alpha\beta}{}_{\lambda}{}_{;\rho}+4(\hat{\xi}\pi'-\hat{\pi}\xi')^{\lambda\rho}g^{\alpha\beta}\phi_{\lambda}{}_{;\rho}+2(\hat{\xi}\pi'-\hat{\pi}\xi')^{\alpha\beta}\phi^{\rho}{}_{;\rho}
\\ &-8(\hat{\xi}\pi'-\hat{\pi}\xi')^{(\alpha|\lambda|}\triangledown^{\beta)}\phi_{\lambda}-8(\hat{\xi}\pi'-\hat{\pi}\xi')^{(\alpha|\lambda|}\triangledown_{\lambda}\phi^{\beta)}+8(\hat{\xi}\pi'-\hat{\pi}\xi')^{(\alpha|\lambda}\triangledown^{\rho|}\phi^{\beta)}{}_{\lambda\rho}
\\ &-8(\hat{\xi}\pi'-\hat{\pi}\xi')^{\lambda\rho}\triangledown^{(\alpha}\phi^{\beta)}{}_{\lambda\rho} \ ,
\end{split}
\end{equation}
where we denoted
\begin{equation}
(\hat{\xi}\pi'-\hat{\pi}\xi')^{\mu\nu} = \hat{\xi}^{\mu \nu}
\pi^\sigma{}_{\sigma} - \hat{\pi}^{\mu \nu} \xi^\sigma{}_{\sigma} \ .
\end{equation}
It might at first seem surprising that the commutator contains traces
of the gauge parameters, whereas in a single gauge transformation only
their traceless part contributes. This is due to the fact that the
notion of the trace is field-dependent (it depends on the metric), and
that the field changes under the gauge transformation. 

Let us briefly explain this phenomenon in a very simple example.
Consider an infinitesimal rotation of a vector $\vec{x} \in \mathbb{R}^3$
parameterised by a vector $\vec{v}\in \mathbb{R}^3$,
\begin{equation}
\delta_{v} \vec{x} = \vec{v} \times \vec{x} \ .
\end{equation}
Obviously the component of $\vec{v}$ parallel to $\vec{x}$,
i.e. $\vec{v}_{\parallel}=\frac{(\vec{v} \cdot \vec{x})}{\| x\|^2}\vec{x}$, 
does not contribute to the rotation. However the commutator of two rotations is given by
\begin{equation}
\left[ \delta_v, \delta_w \right] \vec{x} = (\vec{w} \times \vec{v})
\times \vec{x} = \vec{v} (\vec{w} \cdot \vec{x}) - \vec{w} (\vec{v}
\cdot \vec{x}) \ .
\end{equation}
Therefore the components parallel to $\vec{x}$ contribute in the
commutator although an individual rotation only depends on the
component orthogonal to $\vec{x}$. This is completely analogous to the observation above
that the traces of the spin-3 parameter contribute to the commutator.
\smallskip

This concludes the computation of the commutator of two spin-3 transformations at linear order in the spin-3 field. Together with the expression derived for the commutator of a spin-2 with either a spin-2 or spin-3 transformation, which are exact results, we have therefore determined the gauge algebra to leading order.


\subsection*{Acknowledgements}

We thank Andrea Campoleoni, Teake Nutma and Massimo Taronna for very helpful discussions. In particular we are indebted to Teake Nutma who taught us how to use the package xAct~\cite{xAct}, without which the computations in this paper would not have been possible. 

\appendix 
\section{Conventions}
\label{sec:AppendixConventions}

We denote symmetrisation by a pair of parentheses,
\begin{equation}
A_{(\mu}\, B_{\nu)} \, = \frac{1}{2} \left(A_{\mu}\, B_{\nu} +
A_{\nu}\,B_{\mu} \right) \ .
\end{equation}
Likewise square brackets denote antisymmetrisation. We often omit contracted indices of a tensor to simplify notation, for example  
\begin{equation}
\label{eq:traceOfPhi}
\phi_\mu \equiv \phi_{\mu\lambda}{}^\lambda\ .
\end{equation}
Furthermore we will use hats to denote the traceless projection of a contravariant rank 2 tensor, i.e.
\begin{equation}
\label{eq:tracelessProjOfParameter}
\hat{\xi}^{\mu \nu}=\left(\delta^\mu_\sigma \,\delta^\nu_\kappa -
\tfrac{1}{3} \,g^{\mu\nu}\,g_{\sigma \kappa} \right) \xi^{\sigma \kappa} \ .
\end{equation}
The algebra $sl(3,\mathbb{R})$ can be given in terms of generators
$J_{a}$ and $T_{ab}$ with the commutation relations
\begin{subequations}
\label{sl3}
\begin{align}
& \left[\, J_a \comma J_b \,\right] \, = \, \epsilon_{abc}\, J^c \, , \label{JJ} \\[6pt]
& \left[\, J_a \comma T_{bc} \,\right] \, = \, 2\,\epsilon^d{}_{a(b} T_{c)d} \, , \label{JT} \\[6pt]
& \left[\, T_{ab} \comma T_{cd} \,\right] \, = \, -\, 2 \left(\, \eta_{a(c} \epsilon_{d)be} + \, \eta_{b(c} \epsilon_{d)ae} \,\right) J^e \, , \label{TT}
\end{align}
\end{subequations}
and $T_{[ab]} = \eta^{ab}\, T_{ab}=0$. Here the Levi-Civita symbol is
given by 
\begin{equation}
\epsilon^{012} \, = \, - \ \epsilon_{012} \, = \, 1 \ ,
\end{equation}
and indices can be raised and lowered by
$\eta_{ab}=\text{diag} (-1,1,1)$. 
A $3 \times 3$ matrix representation for the $T_{ab}$ is given by
\begin{equation}
T_{ab} = \left(J_a J_b+J_b J_a-\frac{2}{3}\, \eta_{ab}\, J_c
J^c\right)\ ,
\end{equation}
where $J_a$ is in the three-dimensional representation of $sl(2,\mathbb{R}) \hookrightarrow sl(3,\mathbb{R})$. Furthermore $\{J_{A}\}$ denote a set of five
independent generators built from the matrix representation $T_{ab}$. We use the notation $\{J_{\cA} \}$ for the set of all
generators $\{J_{a},J_{A} \}$.

The Killing form is defined to be one half of the matrix trace in
the fundamental representation of $sl(3,\mathbb{R})$,
\begin{equation}
\kappa_{\cA\cB} = \frac{1}{2}\, \tr \left(J_{\cA}\,J_{\cB} \right)\ ,
\end{equation}
therefore \ $\kappa_{ab}=\eta_{ab}$ and $\kappa_{aB} = 0$. The anti-symmetric and
symmetric structure constants are given by
\begin{align}
f_{\cA\cB\cC} &= \frac{1}{2}\, \tr \left([J_{\cA},J_{\cB}]J_{\cC}\right) \, ,\\
d_{\cA\cB\cC} &= \frac{1}{2}\, \tr \left(\{ J_{\cA},J_{\cB}\}
J_{\cC}\right) \ ,
\end{align}
such that $f_{Abc}=f_{ABC}=0$, $f_{abc}=\epsilon_{abc}$ and $d_{abc}=d_{ABc}=0$.
The structure constants satisfy  a number of identities of which we used
\begin{subequations}
\begin{align}
d_{Abc}\,\kappa^{bc} \,&=\, 0 \ ,\\
d_{Abc}\, d^{A}{}_{de} \,&=\, -\,\tfrac{2}{3}\,\kappa_{bc}\, \kappa_{de} 
+ 2\,\kappa_{d(b}\, \kappa_{c)e} \ .\label{eq:dd-identity}
\end{align}
\end{subequations}

\section{Bianchi-like identity}
\label{app:bianchi}

For the curvature of the spin connection we have the following
Bianchi-like identity,
\begin{equation}\label{app:eq:Bianchi}
f^\mathcal{A}{}_{\mathcal{B}\mathcal{C}} \,R^\mathcal{B}_{[ \mu \nu}\,
e_{\rho]}^\mathcal{C}= 0 \ .
\end{equation}
For convenience we display its proof here. We evaluate
\begin{equation}
\begin{split}
f^\mathcal{A}{}_{\mathcal{B}\mathcal{C}} R^\mathcal{B}_{[ \mu \nu} e_{\rho]}^\mathcal{C} &= f^\mathcal{A}{}_{\mathcal{B}\mathcal{C}} \left( \partial_{[\mu} \omega_\nu^\mathcal{B} e^\mathcal{C}_{\rho]} + \tfrac{1}{2} f^\mathcal{B}{}_{\mathcal{E}\mathcal{F}} \omega^\mathcal{E}_{[\mu} \omega_\nu^\mathcal{F} e^\mathcal{C}_{\rho]} \right) \\
&= f^\mathcal{A}{}_{\mathcal{B}\mathcal{C}} \partial_{[\mu} \omega_\nu^\mathcal{B} e^\mathcal{C}_{\rho]} - \tfrac{1}{2} \left( f^\mathcal{A}{}_{\mathcal{B}\mathcal{E}} f^\mathcal{B}{}_{\mathcal{F}\mathcal{C}} + f^\mathcal{A}{}_{\mathcal{B}\mathcal{F}} f^\mathcal{B}{}_{\mathcal{C}\mathcal{E}} \right)  \omega^\mathcal{E}_{[\mu} \omega_\nu^\mathcal{F} e^\mathcal{C}_{\rho]}  \\
&=f^\mathcal{A}{}_{\mathcal{B}\mathcal{C}} \partial_{[\mu} \omega_\nu^\mathcal{B} e^\mathcal{C}_{\rho]} - \tfrac{1}{2} f^\mathcal{A}{}_{\mathcal{B}\mathcal{E}} \omega^\mathcal{E}_{[\mu} \partial_\nu e^\mathcal{B}_{\rho]} - \tfrac{1}{2} f^\mathcal{A}{}_{\mathcal{B}\mathcal{F}} \omega^\mathcal{F}_{[\nu} \partial_\mu e^\mathcal{B}_{\rho]} \\
&= f^\mathcal{A}{}_{\mathcal{B}\mathcal{C}} \partial_{[\mu} \omega_\nu^\mathcal{B} e_{\rho]}^\mathcal{C} + f^\mathcal{A}{}_{\mathcal{B}\mathcal{E}} \partial_{[\mu} \omega_\nu^\mathcal{E} e_{\rho]}^\mathcal{B} \\
&= f^\mathcal{A}{}_{\mathcal{B}\mathcal{C}} \partial_{[\mu} \left(
\omega^\mathcal{B}_\nu e^\mathcal{C}_{\rho]} \right) \ .
\end{split}
\end{equation}
Here we have used (\ref{eq:torsionConstraint}) to obtain the third line. By using the torsion constraint (\ref{eq:torsionConstraint}) again we yield
\begin{equation}
f^\mathcal{A}{}_{\mathcal{B}\mathcal{C}} \partial_{[\mu} \left(
\omega^\mathcal{B}_\nu e^\mathcal{C}_{\rho]} \right)  = -
\partial_{[\mu} \partial_\nu e^\mathcal{A}_{\rho]} = 0\ ,
\end{equation}
which concludes the proof of~\eqref{app:eq:Bianchi}.

\section{DDI contributions to gauge algebra}
\label{sec:AppendixDDIContrToGaugeAlgebra}
In the following we will summarise the contributions to the parameter $u^{\alpha \beta}$ of the gauge algebra, given in (\ref{eq:gaugeAlgebraParameter}), which vanish due to dimensional dependent identities. These might be helpful in comparing with our results.\\
First we give the term with a derivative acting on the spin-3 field.
\begin{equation}
\begin{split}
D_1^{\alpha \beta}=&\tfrac{1}{2}\bigl((\hat{\xi}\hat{\pi})^{\alpha\beta\chi\delta}\triangledown_{\delta}\phi_{\chi}-(\hat{\xi}\hat{\pi})^{\chi\delta}{}_{\chi}{}^{\epsilon}g^{\alpha\beta}\triangledown_{\epsilon}\phi_{\delta}+(\hat{\xi}\hat{\pi})^{\chi\delta}{}_{\chi}{}^{\epsilon}\triangledown_{\epsilon}\phi^{\alpha\beta}{}_{\delta}-(\hat{\xi}\hat{\pi})^{\alpha\beta\chi\delta}\triangledown_{\epsilon}\phi_{\chi\delta}{}^{\epsilon} \\&
+2(\hat{\xi}\hat{\pi})^{\chi\delta\epsilon\xi}g^{\alpha\beta}\triangledown_{\xi}\phi_{\chi\delta\epsilon}
-(\hat{\xi}\hat{\pi})^{(\alpha|\chi}{}_{\chi}{}^{\delta|}\triangledown^{\beta)}\phi_{\delta}+2(\hat{\xi}\hat{\pi})^{(\alpha|\chi\delta\epsilon|}\triangledown^{\beta)}\phi_{\chi\delta\epsilon}+(\hat{\xi}\hat{\pi})^{(\alpha|\chi}{}_{\chi}{}^{\delta}\triangledown^{\epsilon|}\phi^{\beta)}{}_{\delta\epsilon}\\&
+(\hat{\xi}\hat{\pi})^{(\alpha|\chi\delta\epsilon|}\triangledown_{\chi}\phi^{\beta)}{}_{\delta\epsilon}-3(\hat{\xi}\hat{\pi})^{(\alpha|\chi\delta\epsilon|}\triangledown_{\delta}\phi^{\beta)}{}_{\chi\epsilon}\bigr)\ .
\end{split}
\end{equation}
Furthermore there are four more quantities with a derivative acting on the parameters.
\begin{equation}
\begin{split}
D_2^{\alpha \beta}=&(\nabla\hat{\xi}\hat{\pi})_{\chi}{}^{\delta\epsilon}{}_{\delta\epsilon}\phi^{\chi}g^{\alpha\beta}-(\nabla\hat{\xi}\hat{\pi})^{\chi\delta\epsilon}{}_{\delta\epsilon}\phi^{\alpha\beta}{}_{\chi}-(\nabla\hat{\xi}\hat{\pi})^{\chi\alpha\beta\delta\epsilon}\phi_{\chi\delta\epsilon}-(\nabla\hat{\xi}\hat{\pi})^{\chi\delta\epsilon\alpha\beta}\phi_{\chi\delta\epsilon}-2(\nabla\hat{\xi}\hat{\pi})^{\chi\delta\epsilon}{}_{\delta}{}^{\xi}g^{\alpha\beta}\phi_{\chi\epsilon\xi}\\&
-2(\nabla\hat{\xi}\hat{\pi})^{\chi(\alpha|\delta|\beta)}{}_{\delta}\phi_{\chi}+2(\nabla\hat{\xi}\hat{\pi})^{\chi(\alpha|\delta|\beta)\epsilon}\phi_{\chi\delta\epsilon}+2(\nabla\hat{\xi}\hat{\pi})^{\chi(\alpha|\delta}{}_{\delta}{}^{\epsilon|}\phi^{\beta)}{}_{\chi\epsilon}+2(\nabla\hat{\xi}\hat{\pi})^{\chi\delta\epsilon(\alpha}{}_{\delta}\phi^{\beta)}{}_{\chi\epsilon}
\end{split}
\end{equation}
\begin{equation}
\begin{split}
D_3^{\alpha \beta}=& (\nabla\hat{\xi}\hat{\pi})^{\delta}{}_{\chi\delta}{}^{\alpha\beta}\phi^{\chi}+(\nabla\hat{\xi}\hat{\pi})^{\delta}{}_{\chi}{}^{\epsilon}{}_{\delta\epsilon}\phi^{\chi}g^{\alpha\beta}-(\nabla\hat{\xi}\hat{\pi})^{\chi\delta\epsilon}{}_{\chi\delta}\phi^{\alpha\beta}{}_{\epsilon}-(\nabla\hat{\xi}\hat{\pi})^{\chi\delta\epsilon\alpha\beta}\phi_{\chi\delta\epsilon}-(\nabla\hat{\xi}\hat{\pi})^{\chi\delta\epsilon}{}_{\delta}{}^{\xi}g^{\alpha\beta}\phi_{\chi\epsilon\xi} \\&
+(\nabla\hat{\xi}\hat{\pi})^{\chi}{}_{\chi}{}^{\delta\epsilon\xi}g^{\alpha\beta}\phi_{\delta\epsilon\xi}-(\nabla\hat{\xi}\hat{\pi})^{\chi\delta\epsilon}{}_{\chi}{}^{\xi}g^{\alpha\beta}\phi_{\delta\epsilon\xi}-(\nabla\hat{\xi}\hat{\pi})^{(\alpha|\chi\delta|\beta)}{}_{\chi}\phi_{\delta}-(\nabla\hat{\xi}\hat{\pi})^{\chi(\alpha|\delta|\beta)}{}_{\chi}\phi_{\delta} \\&
-(\nabla\hat{\xi}\hat{\pi})^{(\alpha\beta)\chi\delta\epsilon}\phi_{\chi\delta\epsilon} 
+(\nabla\hat{\xi}\hat{\pi})^{(\alpha|\chi\delta|\beta)\epsilon}\phi_{\chi\delta\epsilon}+(\nabla\hat{\xi}\hat{\pi})^{(\alpha|\chi\delta}{}_{\chi}{}^{\epsilon|}\phi^{\beta)}{}_{\delta\epsilon}+(\nabla\hat{\xi}\hat{\pi})^{\chi(\alpha|\delta|\beta)\epsilon}\phi_{\chi\delta\epsilon} \\&
+(\nabla\hat{\xi}\hat{\pi})^{\chi(\alpha|\delta}{}_{\chi}{}^{\epsilon|}\phi^{\beta)}{}_{\delta\epsilon} -2(\nabla\hat{\xi}\hat{\pi})^{\chi}{}_{\chi}{}^{\delta(\alpha|\epsilon|}\phi^{\beta)}{}_{\delta\epsilon}+(\nabla\hat{\xi}\hat{\pi})^{\chi\delta\epsilon(\alpha}{}_{\chi}\phi^{\beta)}{}_{\delta\epsilon}+(\nabla\hat{\xi}\hat{\pi})^{\chi\delta\epsilon(\alpha}{}_{\delta}\phi^{\beta)}{}_{\chi\epsilon}
\end{split}
\end{equation}
\begin{equation}
\begin{split}
D_4^{\alpha \beta}=&-(\nabla\hat{\xi}\hat{\pi})^{\delta\alpha\beta}{}_{\chi\delta}\phi^{\chi}-(\nabla\hat{\xi}\hat{\pi})^{\delta}{}_{\delta}{}^{\epsilon}{}_{\chi\epsilon}\phi^{\chi}g^{\alpha\beta}+(\nabla\hat{\xi}\hat{\pi})^{\chi}{}_{\chi}{}^{\delta}{}_{\delta}{}^{\epsilon}\phi^{\alpha\beta}{}_{\epsilon}+(\nabla\hat{\xi}\hat{\pi})^{\chi\alpha\beta\delta\epsilon}\phi_{\chi\delta\epsilon}+(\nabla\hat{\xi}\hat{\pi})^{\chi\delta\epsilon}{}_{\delta}{}^{\xi}g^{\alpha\beta}\phi_{\chi\epsilon\xi} \\&
+(\nabla\hat{\xi}\hat{\pi})^{\chi}{}_{\chi}{}^{\delta\epsilon\xi}g^{\alpha\beta}\phi_{\delta\epsilon\xi}-(\nabla\hat{\xi}\hat{\pi})^{\chi\delta\epsilon}{}_{\chi}{}^{\xi}g^{\alpha\beta}\phi_{\delta\epsilon\xi}+(\nabla\hat{\xi}\hat{\pi})^{(\alpha\beta)\chi}{}_{\chi}{}^{\delta}\phi_{\delta}+(\nabla\hat{\xi}\hat{\pi})^{\chi(\alpha}{}_{\chi}{}^{\beta)\delta}\phi_{\delta} \\&
-(\nabla\hat{\xi}\hat{\pi})^{(\alpha\beta)\chi\delta\epsilon}\phi_{\chi\delta\epsilon}+(\nabla\hat{\xi}\hat{\pi})^{(\alpha|\chi\delta|\beta)\epsilon}\phi_{\chi\delta\epsilon} 
-(\nabla\hat{\xi}\hat{\pi})^{(\alpha|\chi\delta}{}_{\chi}{}^{\epsilon|}\phi^{\beta)}{}_{\delta\epsilon}-(\nabla\hat{\xi}\hat{\pi})^{\chi(\alpha|\delta|\beta)\epsilon}\phi_{\chi\delta\epsilon} \\&
-(\nabla\hat{\xi}\hat{\pi})^{\chi(\alpha}{}_{\chi}{}^{|\delta\epsilon|}\phi^{\beta)}{}_{\delta\epsilon}+2(\nabla\hat{\xi}\hat{\pi})^{\chi(\alpha|\delta}{}_{\chi}{}^{\epsilon|}\phi^{\beta)}{}_{\delta\epsilon} 
-(\nabla\hat{\xi}\hat{\pi})^{\chi(\alpha|\delta}{}_{\delta}{}^{\epsilon|}\phi^{\beta)}{}_{\chi\epsilon}-(\nabla\hat{\xi}\hat{\pi})^{\chi}{}_{\chi}{}^{\delta(\alpha|\epsilon|}\phi^{\beta)}{}_{\delta\epsilon}
\end{split}
\end{equation}
\begin{equation}
\begin{split}
D_5^{\alpha \beta}=&-(\nabla\hat{\xi}\hat{\pi})^{\delta\alpha\beta}{}_{\chi\delta}\phi^{\chi}-(\nabla\hat{\xi}\hat{\pi})^{\delta}{}_{\chi\delta}{}^{\alpha\beta}\phi^{\chi}+(\nabla\hat{\xi}\hat{\pi})_{\chi}{}^{\delta\epsilon}{}_{\delta\epsilon}\phi^{\chi}g^{\alpha\beta}-(\nabla\hat{\xi}\hat{\pi})^{\delta}{}_{\chi}{}^{\epsilon}{}_{\delta\epsilon}\phi^{\chi}g^{\alpha\beta}-(\nabla\hat{\xi}\hat{\pi})^{\delta}{}_{\delta}{}^{\epsilon}{}_{\chi\epsilon}\phi^{\chi}g^{\alpha\beta}\\& 
+(\nabla\hat{\xi}\hat{\pi})^{(\alpha\beta)\chi}{}_{\chi}{}^{\delta}\phi_{\delta} +(\nabla\hat{\xi}\hat{\pi})^{(\alpha|\chi\delta|\beta)}{}_{\chi}\phi_{\delta}-(\nabla\hat{\xi}\hat{\pi})^{(\alpha|\chi\delta|}{}_{\chi\delta}\phi^{\beta)}+(\nabla\hat{\xi}\hat{\pi})^{\chi(\alpha}{}_{\chi}{}^{\beta)\delta}\phi_{\delta}+(\nabla\hat{\xi}\hat{\pi})^{\chi(\alpha|\delta|\beta)}{}_{\chi}\phi_{\delta} \\&
-2(\nabla\hat{\xi}\hat{\pi})^{\chi(\alpha|\delta|\beta)}{}_{\delta}\phi_{\chi}+(\nabla\hat{\xi}\hat{\pi})^{\chi(\alpha|\delta|}{}_{\chi\delta}\phi^{\beta)}+(\nabla\hat{\xi}\hat{\pi})^{\chi}{}_{\chi}{}^{\delta(\alpha}{}_{\delta}\phi^{\beta)}\ .
\end{split}
\end{equation}
The quantities $(\nabla\hat{\xi}\hat{\pi})$ and $(\hat{\xi}\hat{\pi})$ are defined in (\ref{eq:commutatorNablaXiPiParameter}) and (\ref{eq:commutatorXiPiParameter}).

\section{Higher order corrections to spin-3 transformations}
\label{sec:AppendixGaugeTransformations}
In this section we list the higher order corrections to the gauge transformations given in section \ref{sec:gaugeTransformations}. By making a particular choice of the undetermined constants due to dimensional dependent identities we reduced the size of the expressions considerably. For the gauge transformation of the spin-3 field there are corrections with a derivative acting on the $\phi$ field,
\begin{align}
&(\hat{\xi} \phi \nabla \phi)_{\alpha \beta \chi} =
18\Bigl(2\phi^{\delta}\hat{\xi}_{\delta}{}^{\epsilon}\triangledown_{\epsilon}\phi_{\alpha\beta\chi}+\hat{\xi}^{\delta\epsilon}\bigl(3\phi_{\delta\epsilon}{}^{\xi}\triangledown_{\xi}\phi_{\alpha\beta\chi}+\phi_{\alpha\beta\chi}(7\triangledown_{\epsilon}\phi_{\delta}-3\triangledown_{\xi}\phi_{\delta\epsilon}{}^{\xi})\bigr)-4\phi_{(\alpha}\hat{\xi}_{\beta\chi)}\triangledown^{\delta}\phi_{\delta}\nonumber\\&+8\phi_{(\alpha}\hat{\xi}_{\beta}{}^{\delta}\triangledown_{\chi)}\phi_{\delta}+9\phi_{(\alpha}\hat{\xi}_{\beta}{}^{\delta}\triangledown_{|\delta|}\phi_{\chi)}-5\phi^{\delta}\hat{\xi}_{(\alpha\beta}\triangledown_{|\delta|}\phi_{\chi)}-2\phi^{\delta}\hat{\xi}_{(\alpha|\delta|}\triangledown_{\beta}\phi_{\chi)}-8\phi_{(\alpha}\hat{\xi}_{\beta}{}^{\delta}\triangledown^{\epsilon}\phi_{\chi)\delta\epsilon}\nonumber\\&-2\phi_{(\alpha}\hat{\xi}^{\delta\epsilon}\triangledown_{|\delta|}\phi_{\beta\chi)\epsilon}+\phi^{\delta}\hat{\xi}_{(\alpha|\delta|}\triangledown^{\epsilon}\phi_{\beta\chi)\epsilon}+13\phi^{\delta}\hat{\xi}_{(\alpha}{}^{\epsilon}\triangledown_{|\delta|}\phi_{\beta\chi)\epsilon}-3\phi^{\delta}\hat{\xi}_{(\alpha}{}^{\epsilon}\triangledown_{|\epsilon|}\phi_{\beta\chi)\delta}+12\hat{\xi}_{(\alpha\beta}\phi_{\chi)}{}^{\delta\epsilon}\triangledown_{\delta}\phi_{\epsilon}\nonumber\\&+5\hat{\xi}_{(\alpha}{}^{\delta}\phi_{\beta\chi)\delta}\triangledown^{\epsilon}\phi_{\epsilon}-12\hat{\xi}_{(\alpha}{}^{\delta}\phi_{\beta\chi)}{}^{\epsilon}\triangledown_{\delta}\phi_{\epsilon}-7\hat{\xi}_{(\alpha}{}^{\delta}\phi_{\beta\chi)}{}^{\epsilon}\triangledown_{\epsilon}\phi_{\delta}-6\hat{\xi}_{(\alpha}{}^{\delta}\phi_{\beta|\delta|}{}^{\epsilon}\triangledown_{\chi)}\phi_{\epsilon}\nonumber\\&+8\hat{\xi}_{(\alpha}{}^{\delta}\phi_{\beta|\delta}{}^{\epsilon}\triangledown_{\epsilon|}\phi_{\chi)}-3\hat{\xi}^{\delta\epsilon}\phi_{(\alpha\beta|\delta|}\triangledown_{\chi)}\phi_{\epsilon}-4\hat{\xi}^{\delta\epsilon}\phi_{(\alpha\beta|\delta}\triangledown_{\epsilon|}\phi_{\chi)}-8\hat{\xi}_{(\alpha\beta}\phi_{\chi)}{}^{\delta\epsilon}\triangledown^{\xi}\phi_{\delta\epsilon\xi}+\hat{\xi}_{(\alpha\beta}\phi^{\delta\epsilon\xi}\triangledown_{\chi)}\phi_{\delta\epsilon\xi}\nonumber\\&+5\hat{\xi}_{(\alpha\beta}\phi^{\delta\epsilon\xi}\triangledown_{|\delta|}\phi_{\chi)\epsilon\xi}+11\hat{\xi}_{(\alpha}{}^{\delta}\phi_{\beta\chi)}{}^{\epsilon}\triangledown^{\xi}\phi_{\delta\epsilon\xi}+\hat{\xi}_{(\alpha}{}^{\delta}\phi_{\beta}{}^{\epsilon\xi}\triangledown_{\chi)}\phi_{\delta\epsilon\xi}+6\hat{\xi}_{(\alpha}{}^{\delta}\phi_{\beta|\delta|}{}^{\epsilon}\triangledown^{\xi}\phi_{\chi)\epsilon\xi}\nonumber\\&+3\hat{\xi}_{(\alpha}{}^{\delta}\phi_{\beta}{}^{\epsilon\xi}\triangledown_{|\delta|}\phi_{\chi)\epsilon\xi}-18\hat{\xi}_{(\alpha}{}^{\delta}\phi_{\beta}{}^{\epsilon\xi}\triangledown_{|\epsilon|}\phi_{\chi)\delta\xi}-\hat{\xi}_{(\alpha}{}^{\delta}\phi_{|\delta|}{}^{\epsilon\xi}\triangledown_{\beta}\phi_{\chi)\epsilon\xi}-11\hat{\xi}_{(\alpha}{}^{\delta}\phi_{|\delta}{}^{\epsilon\xi}\triangledown_{\epsilon|}\phi_{\beta\chi)\xi}\nonumber\\&-9\hat{\xi}^{\delta\epsilon}\phi_{(\alpha\beta}{}^{\xi}\triangledown_{\chi)}\phi_{\delta\epsilon\xi}+3\hat{\xi}^{\delta\epsilon}\phi_{(\alpha\beta|\delta|}\triangledown^{\xi}\phi_{\chi)\epsilon\xi}+7\hat{\xi}^{\delta\epsilon}\phi_{(\alpha\beta}{}^{\xi}\triangledown_{|\delta|}\phi_{\chi)\epsilon\xi}+3\hat{\xi}^{\delta\epsilon}\phi_{(\alpha\beta}{}^{\xi}\triangledown_{|\xi|}\phi_{\chi)\delta\epsilon}\nonumber\\&-7\hat{\xi}^{\delta\epsilon}\phi_{(\alpha|\delta}{}^{\xi}\triangledown_{\epsilon|}\phi_{\beta\chi)\xi}-9\hat{\xi}^{\delta\epsilon}\phi_{(\alpha|\delta}{}^{\xi}\triangledown_{\xi|}\phi_{\beta\chi)\epsilon}-3\phi_{(\alpha}\hat{\xi}^{\delta\epsilon}g_{\beta\chi)}\triangledown_{\delta}\phi_{\epsilon}+2\phi^{\delta}\hat{\xi}_{(\alpha|\delta|}g_{\beta\chi)}\triangledown^{\epsilon}\phi_{\epsilon}\nonumber\\&-2\phi^{\delta}\hat{\xi}_{(\alpha}{}^{\epsilon}g_{\beta\chi)}\triangledown_{\delta}\phi_{\epsilon}-3\phi^{\delta}\hat{\xi}_{(\alpha}{}^{\epsilon}g_{\beta\chi)}\triangledown_{\epsilon}\phi_{\delta}-3\phi^{\delta}\hat{\xi}_{\delta}{}^{\epsilon}g_{(\alpha\beta}\triangledown_{\chi)}\phi_{\epsilon}-3\phi^{\delta}\hat{\xi}_{\delta}{}^{\epsilon}g_{(\alpha\beta}\triangledown_{|\epsilon|}\phi_{\chi)}\nonumber\\&+\phi_{(\alpha}\hat{\xi}^{\delta\epsilon}g_{\beta\chi)}\triangledown^{\xi}\phi_{\delta\epsilon\xi}+3\phi^{\delta}\hat{\xi}_{(\alpha}{}^{\epsilon}g_{\beta\chi)}\triangledown^{\xi}\phi_{\delta\epsilon\xi}+3\phi^{\delta}\hat{\xi}_{\delta}{}^{\epsilon}g_{(\alpha\beta}\triangledown^{\xi}\phi_{\chi)\epsilon\xi}-7\phi^{\delta}\hat{\xi}^{\epsilon\xi}g_{(\alpha\beta}\triangledown_{|\delta|}\phi_{\chi)\epsilon\xi}\nonumber\\&+3\phi^{\delta}\hat{\xi}^{\epsilon\xi}g_{(\alpha\beta}\triangledown_{|\epsilon|}\phi_{\chi)\delta\xi}-3\hat{\xi}_{(\alpha}{}^{\delta}g_{\beta\chi)}\phi_{\delta}{}^{\epsilon\xi}\triangledown_{\epsilon}\phi_{\xi}-2\hat{\xi}^{\delta\epsilon}g_{(\alpha\beta}\phi_{\chi)\delta\epsilon}\triangledown^{\xi}\phi_{\xi}+9\hat{\xi}^{\delta\epsilon}g_{(\alpha\beta}\phi_{\chi)\delta}{}^{\xi}\triangledown_{\epsilon}\phi_{\xi}\nonumber\\&+3\hat{\xi}^{\delta\epsilon}g_{(\alpha\beta}\phi_{\chi)\delta}{}^{\xi}\triangledown_{\xi}\phi_{\epsilon}-4\hat{\xi}^{\delta\epsilon}g_{(\alpha\beta}\phi_{|\delta\epsilon}{}^{\xi}\triangledown_{\xi|}\phi_{\chi)}+\hat{\xi}_{(\alpha}{}^{\delta}g_{\beta\chi)}\phi_{\delta}{}^{\epsilon\xi}\triangledown^{\gamma}\phi_{\epsilon\xi\gamma}+\hat{\xi}_{(\alpha}{}^{\delta}g_{\beta\chi)}\phi^{\epsilon\xi\gamma}\triangledown_{\delta}\phi_{\epsilon\xi\gamma}\nonumber\\&-\hat{\xi}_{(\alpha}{}^{\delta}g_{\beta\chi)}\phi^{\epsilon\xi\gamma}\triangledown_{\epsilon}\phi_{\delta\xi\gamma}-8\hat{\xi}^{\delta\epsilon}g_{(\alpha\beta}\phi_{\chi)\delta}{}^{\xi}\triangledown^{\gamma}\phi_{\epsilon\xi\gamma}-6\hat{\xi}^{\delta\epsilon}g_{(\alpha\beta}\phi_{\chi)}{}^{\xi\gamma}\triangledown_{\delta}\phi_{\epsilon\xi\gamma}+6\hat{\xi}^{\delta\epsilon}g_{(\alpha\beta}\phi_{\chi)}{}^{\xi\gamma}\triangledown_{\xi}\phi_{\delta\epsilon\gamma}\nonumber\\&+6\hat{\xi}^{\delta\epsilon}g_{(\alpha\beta}\phi_{|\delta|}{}^{\xi\gamma}\triangledown_{\chi)}\phi_{\epsilon\xi\gamma}+8\hat{\xi}^{\delta\epsilon}g_{(\alpha\beta}\phi_{|\delta}{}^{\xi\gamma}\triangledown_{\xi|}\phi_{\chi)\epsilon\gamma}\Bigr)
\ .
\end{align}
Then there are contributions with a derivative acting on the parameter.
\begin{align}
&(\nabla \hat{\xi} \phi \phi)_{\alpha \beta \chi}
=-9\Bigl(12\phi^{\delta}\phi_{\alpha\beta\chi}\triangledown_{\epsilon}\hat{\xi}_{\delta}{}^{\epsilon}+4\phi_{\alpha\beta\chi}\phi^{\delta\epsilon\xi}\triangledown_{\xi}\hat{\xi}_{\delta\epsilon}+14\phi_{(\alpha}\phi_{\beta}\triangledown^{\delta}\hat{\xi}_{\chi)\delta}+6\phi_{(\alpha}\phi^{\delta}\triangledown_{\beta}\hat{\xi}_{\chi)\delta}\nonumber\\&+\phi^{\delta}\phi_{\delta}\triangledown_{(\alpha}\hat{\xi}_{\beta\chi)}-16\phi_{(\alpha}\phi_{\beta\chi)}{}^{\delta}\triangledown^{\epsilon}\hat{\xi}_{\delta\epsilon}+16\phi_{(\alpha}\phi_{\beta}{}^{\delta\epsilon}\triangledown_{\chi)}\hat{\xi}_{\delta\epsilon}-8\phi_{(\alpha}\phi_{\beta}{}^{\delta\epsilon}\triangledown_{|\delta|}\hat{\xi}_{\chi)\epsilon}-12\phi^{\delta}\phi_{(\alpha\beta}{}^{\epsilon}\triangledown_{\chi)}\hat{\xi}_{\delta\epsilon}\nonumber\\&-8\phi^{\delta}\phi_{(\alpha\beta|\delta|}\triangledown^{\epsilon}\hat{\xi}_{\chi)\epsilon}+32\phi^{\delta}\phi_{(\alpha\beta}{}^{\epsilon}\triangledown_{|\delta|}\hat{\xi}_{\chi)\epsilon}-12\phi^{\delta}\phi_{(\alpha\beta}{}^{\epsilon}\triangledown_{|\epsilon|}\hat{\xi}_{\chi)\delta}+8\phi^{\delta}\phi_{(\alpha|\delta}{}^{\epsilon}\triangledown_{\epsilon|}\hat{\xi}_{\beta\chi)}\nonumber\\&+4\phi_{(\alpha\beta}{}^{\delta}\phi_{\chi)\delta}{}^{\epsilon}\triangledown^{\xi}\hat{\xi}_{\epsilon\xi}-4\phi_{(\alpha\beta}{}^{\delta}\phi_{\chi)}{}^{\epsilon\xi}\triangledown_{\epsilon}\hat{\xi}_{\delta\xi}+14\phi_{(\alpha\beta}{}^{\delta}\phi_{|\delta|}{}^{\epsilon\xi}\triangledown_{\chi)}\hat{\xi}_{\epsilon\xi}-12\phi_{(\alpha\beta}{}^{\delta}\phi_{|\delta}{}^{\epsilon\xi}\triangledown_{\epsilon|}\hat{\xi}_{\chi)\xi}\nonumber\\&-6\phi_{(\alpha}{}^{\delta\epsilon}\phi_{\beta|\delta\epsilon|}\triangledown^{\xi}\hat{\xi}_{\chi)\xi}+8\phi_{(\alpha}{}^{\delta\epsilon}\phi_{|\delta\epsilon}{}^{\xi}\triangledown_{\xi|}\hat{\xi}_{\beta\chi)}-12\phi_{(\alpha}\phi^{\delta}g_{\beta\chi)}\triangledown^{\epsilon}\hat{\xi}_{\delta\epsilon}+3\phi^{\delta}\phi^{\epsilon}g_{(\alpha\beta}\triangledown_{\chi)}\hat{\xi}_{\delta\epsilon}\nonumber\\&-\phi^{\delta}\phi_{\delta}g_{(\alpha\beta}\triangledown^{\epsilon}\hat{\xi}_{\chi)\epsilon}-4\phi_{(\alpha}g_{\beta\chi)}\phi^{\delta\epsilon\xi}\triangledown_{\delta}\hat{\xi}_{\epsilon\xi}+10\phi^{\delta}g_{(\alpha\beta}\phi_{\chi)\delta}{}^{\epsilon}\triangledown^{\xi}\hat{\xi}_{\epsilon\xi}-16\phi^{\delta}g_{(\alpha\beta}\phi_{\chi)}{}^{\epsilon\xi}\triangledown_{\delta}\hat{\xi}_{\epsilon\xi}\nonumber\\&+12\phi^{\delta}g_{(\alpha\beta}\phi_{\chi)}{}^{\epsilon\xi}\triangledown_{\epsilon}\hat{\xi}_{\delta\xi}-6\phi^{\delta}g_{(\alpha\beta}\phi_{|\delta|}{}^{\epsilon\xi}\triangledown_{\chi)}\hat{\xi}_{\epsilon\xi}+2\phi^{\delta}g_{(\alpha\beta}\phi_{|\delta}{}^{\epsilon\xi}\triangledown_{\epsilon|}\hat{\xi}_{\chi)\xi}+2g_{(\alpha\beta}\phi_{\chi)}{}^{\delta\epsilon}\phi_{\delta\epsilon}{}^{\xi}\triangledown^{\gamma}\hat{\xi}_{\xi\gamma}\nonumber\\&-2g_{(\alpha\beta}\phi_{\chi)}{}^{\delta\epsilon}\phi_{\delta}{}^{\xi\gamma}\triangledown_{\epsilon}\hat{\xi}_{\xi\gamma}+10g_{(\alpha\beta}\phi_{\chi)}{}^{\delta\epsilon}\phi_{\delta}{}^{\xi\gamma}\triangledown_{\xi}\hat{\xi}_{\epsilon\gamma}+g_{(\alpha\beta}\phi^{\delta\epsilon\xi}\phi_{|\delta\epsilon\xi|}\triangledown^{\gamma}\hat{\xi}_{\chi)\gamma}\nonumber\\&-2g_{(\alpha\beta}\phi^{\delta\epsilon\xi}\phi_{|\delta\epsilon}{}^{\gamma}\triangledown_{\xi|}\hat{\xi}_{\chi)\gamma}-16\phi_{(\alpha}\phi^{\delta}\triangledown_{|\delta|}\hat{\xi}_{\beta\chi)}\Bigr)\ .
\end{align}
Finally the transformation of the metric to cubic order is given by
\begin{align}
&(\hat{\xi} \phi \phi \nabla \phi)_{\alpha \beta}=18\bigl(16\phi^{\chi}\hat{\xi}^{\delta\epsilon}\phi_{\alpha\beta\delta}\triangledown_{\chi}\phi_{\epsilon}-8\phi^{\chi}\hat{\xi}^{\delta\epsilon}g_{\alpha\beta}\phi_{\delta\epsilon}{}^{\xi}\triangledown_{\chi}\phi_{\xi}+16\phi^{\chi}\hat{\xi}^{\delta\epsilon}\phi_{\delta\epsilon}{}^{\xi}\triangledown_{\chi}\phi_{\alpha\beta\xi}\nonumber\\&+16\phi^{\chi}\hat{\xi}^{\delta\epsilon}g_{\alpha\beta}\phi_{\delta}{}^{\xi\gamma}\triangledown_{\chi}\phi_{\epsilon\xi\gamma}+5\phi^{\chi}\phi^{\delta}\hat{\xi}_{\alpha\beta}\triangledown_{\delta}\phi_{\chi}-14\phi_{\chi}\phi^{\chi}\hat{\xi}_{\alpha\beta}\triangledown_{\delta}\phi^{\delta}+4\phi^{\chi}\phi^{\delta}\hat{\xi}_{\chi}{}^{\epsilon}g_{\alpha\beta}\triangledown_{\delta}\phi_{\epsilon}\nonumber\\&+12\hat{\xi}^{\chi\delta}\phi_{\alpha\beta}{}^{\epsilon}\phi_{\chi\epsilon}{}^{\xi}\triangledown_{\delta}\phi_{\xi}+8\hat{\xi}^{\chi\delta}g_{\alpha\beta}\phi_{\chi}{}^{\epsilon\xi}\phi_{\epsilon\xi}{}^{\gamma}\triangledown_{\delta}\phi_{\gamma}-10\phi^{\chi}\phi^{\delta}\hat{\xi}^{\epsilon\xi}g_{\alpha\beta}\triangledown_{\delta}\phi_{\chi\epsilon\xi}-8\phi^{\chi}\hat{\xi}_{\chi}{}^{\delta}g_{\alpha\beta}\phi^{\epsilon\xi\gamma}\triangledown_{\delta}\phi_{\epsilon\xi\gamma}\nonumber\\&+8\hat{\xi}^{\chi\delta}g_{\alpha\beta}\phi_{\chi}{}^{\epsilon\xi}\phi_{\epsilon}{}^{\gamma\eta}\triangledown_{\delta}\phi_{\xi\gamma\eta}+12\phi^{\chi}\phi^{\delta}\hat{\xi}_{\chi}{}^{\epsilon}g_{\alpha\beta}\triangledown_{\epsilon}\phi_{\delta}-20\phi_{\chi}\phi^{\chi}\hat{\xi}^{\delta\epsilon}g_{\alpha\beta}\triangledown_{\epsilon}\phi_{\delta}+14\phi^{\chi}\hat{\xi}_{\alpha\beta}\phi_{\chi}{}^{\delta\epsilon}\triangledown_{\epsilon}\phi_{\delta}\nonumber\\&-\phi^{\chi}\phi^{\delta}\hat{\xi}_{\chi\delta}g_{\alpha\beta}\triangledown_{\epsilon}\phi^{\epsilon}-9\phi^{\chi}\phi^{\delta}\hat{\xi}_{\alpha\beta}\triangledown_{\epsilon}\phi_{\chi\delta}{}^{\epsilon}+20\phi^{\chi}\hat{\xi}^{\delta\epsilon}g_{\alpha\beta}\phi_{\chi}{}^{\xi\gamma}\triangledown_{\epsilon}\phi_{\delta\xi\gamma}+8\hat{\xi}^{\chi\delta}\phi_{\alpha\beta}{}^{\epsilon}\phi_{\chi\epsilon}{}^{\xi}\triangledown_{\xi}\phi_{\delta}\nonumber\\&-8\hat{\xi}^{\chi\delta}\phi_{\alpha\beta}{}^{\epsilon}\phi_{\chi\delta}{}^{\xi}\triangledown_{\xi}\phi_{\epsilon}+18\hat{\xi}_{\alpha\beta}\phi_{\chi\delta}{}^{\xi}\phi^{\chi\delta\epsilon}\triangledown_{\xi}\phi_{\epsilon}-10\phi^{\chi}\hat{\xi}^{\delta\epsilon}g_{\alpha\beta}\phi_{\chi\delta\epsilon}\triangledown_{\xi}\phi^{\xi}+4\hat{\xi}_{\alpha\beta}\phi_{\chi\delta\epsilon}\phi^{\chi\delta\epsilon}\triangledown_{\xi}\phi^{\xi}\nonumber\\&-8\phi^{\chi}\hat{\xi}_{\chi}{}^{\delta}\phi_{\delta}{}^{\epsilon\xi}\triangledown_{\xi}\phi_{\alpha\beta\epsilon}-12\phi^{\chi}\phi^{\delta}\hat{\xi}_{\chi}{}^{\epsilon}g_{\alpha\beta}\triangledown_{\xi}\phi_{\delta\epsilon}{}^{\xi}+10\phi_{\chi}\phi^{\chi}\hat{\xi}^{\delta\epsilon}g_{\alpha\beta}\triangledown_{\xi}\phi_{\delta\epsilon}{}^{\xi}+18\phi^{\chi}\hat{\xi}_{\alpha\beta}\phi_{\chi}{}^{\delta\epsilon}\triangledown_{\xi}\phi_{\delta\epsilon}{}^{\xi}\nonumber\\&+16\hat{\xi}^{\chi\delta}g_{\alpha\beta}\phi_{\chi}{}^{\epsilon\xi}\phi_{\epsilon\xi}{}^{\gamma}\triangledown_{\gamma}\phi_{\delta}+20\hat{\xi}^{\chi\delta}g_{\alpha\beta}\phi_{\chi}{}^{\epsilon\xi}\phi_{\delta\epsilon}{}^{\gamma}\triangledown_{\gamma}\phi_{\xi}+6\hat{\xi}^{\chi\delta}g_{\alpha\beta}\phi_{\chi}{}^{\epsilon\xi}\phi_{\delta\epsilon\xi}\triangledown_{\gamma}\phi^{\gamma}\nonumber\\&+4\hat{\xi}^{\chi\delta}\phi_{\chi}{}^{\epsilon\xi}\phi_{\delta\epsilon}{}^{\gamma}\triangledown_{\gamma}\phi_{\alpha\beta\xi}-24\hat{\xi}^{\chi\delta}\phi_{\chi\delta}{}^{\epsilon}\phi_{\epsilon}{}^{\xi\gamma}\triangledown_{\gamma}\phi_{\alpha\beta\xi}-8\hat{\xi}^{\chi\delta}\phi_{\alpha\beta}{}^{\epsilon}\phi_{\epsilon}{}^{\xi\gamma}\triangledown_{\gamma}\phi_{\chi\delta\xi}+16\phi^{\chi}\hat{\xi}^{\delta\epsilon}g_{\alpha\beta}\phi_{\chi}{}^{\xi\gamma}\triangledown_{\gamma}\phi_{\delta\epsilon\xi}\nonumber\\&+24\hat{\xi}^{\chi\delta}\phi_{\alpha\beta}{}^{\epsilon}\phi_{\chi}{}^{\xi\gamma}\triangledown_{\gamma}\phi_{\delta\epsilon\xi}-12\hat{\xi}_{\alpha\beta}\phi_{\chi}{}^{\xi\gamma}\phi^{\chi\delta\epsilon}\triangledown_{\gamma}\phi_{\delta\epsilon\xi}-16\hat{\xi}^{\chi\delta}\phi_{\alpha\beta\chi}\phi^{\epsilon\xi\gamma}\triangledown_{\gamma}\phi_{\delta\epsilon\xi}-12\hat{\xi}^{\chi\delta}\phi_{\alpha\beta}{}^{\epsilon}\phi_{\chi\epsilon}{}^{\xi}\triangledown_{\gamma}\phi_{\delta\xi}{}^{\gamma}\nonumber\\&-14\hat{\xi}_{\alpha\beta}\phi_{\chi\delta}{}^{\xi}\phi^{\chi\delta\epsilon}\triangledown_{\gamma}\phi_{\epsilon\xi}{}^{\gamma}+8\phi^{\chi}\hat{\xi}_{\chi}{}^{\delta}g_{\alpha\beta}\phi_{\delta}{}^{\epsilon\xi}\triangledown_{\gamma}\phi_{\epsilon\xi}{}^{\gamma}-4\hat{\xi}^{\chi\delta}g_{\alpha\beta}\phi_{\epsilon\xi}{}^{\eta}\phi^{\epsilon\xi\gamma}\triangledown_{\eta}\phi_{\chi\delta\gamma}\nonumber\\&-40\hat{\xi}^{\chi\delta}g_{\alpha\beta}\phi_{\chi}{}^{\epsilon\xi}\phi_{\epsilon}{}^{\gamma\eta}\triangledown_{\eta}\phi_{\delta\xi\gamma}-8\hat{\xi}^{\chi\delta}g_{\alpha\beta}\phi_{\chi}{}^{\epsilon\xi}\phi_{\epsilon\xi}{}^{\gamma}\triangledown_{\eta}\phi_{\delta\gamma}{}^{\eta}-16\hat{\xi}^{\chi\delta}g_{\alpha\beta}\phi_{\chi}{}^{\epsilon\xi}\phi_{\delta\epsilon}{}^{\gamma}\triangledown_{\eta}\phi_{\xi\gamma}{}^{\eta}\nonumber\\&+8\hat{\xi}^{\chi\delta}g_{\alpha\beta}\phi_{\chi\delta}{}^{\epsilon}\phi_{\epsilon}{}^{\xi\gamma}\triangledown_{\eta}\phi_{\xi\gamma}{}^{\eta}+2\phi_{(\alpha}\phi^{\chi}\hat{\xi}_{\beta)\chi}\triangledown^{\delta}\phi_{\delta}-20\phi_{(\alpha}\phi^{\chi}\hat{\xi}_{\beta)}{}^{\delta}\triangledown_{\chi}\phi_{\delta}-12\phi_{(\alpha}\phi^{\chi}\hat{\xi}_{\beta)}{}^{\delta}\triangledown_{\delta}\phi_{\chi}\nonumber\\&+20\phi^{\chi}\phi_{\chi}\hat{\xi}_{(\alpha}{}^{\delta}\triangledown_{\beta)}\phi_{\delta}-12\phi^{\chi}\phi^{\delta}\hat{\xi}_{(\alpha|\chi|}\triangledown_{\beta)}\phi_{\delta}+20\phi^{\chi}\phi_{\chi}\hat{\xi}_{(\alpha}{}^{\delta}\triangledown_{|\delta|}\phi_{\beta)}-4\phi^{\chi}\phi^{\delta}\hat{\xi}_{(\alpha|\chi}\triangledown_{\delta|}\phi_{\beta)}\nonumber\\&+12\phi_{(\alpha}\phi^{\chi}\hat{\xi}_{\beta)}{}^{\delta}\triangledown^{\epsilon}\phi_{\chi\delta\epsilon}+16\phi_{(\alpha}\phi^{\chi}\hat{\xi}^{\delta\epsilon}\triangledown_{|\chi|}\phi_{\beta)\delta\epsilon}-20\phi^{\chi}\phi_{\chi}\hat{\xi}_{(\alpha}{}^{\delta}\triangledown^{\epsilon}\phi_{\beta)\delta\epsilon}+12\phi^{\chi}\phi^{\delta}\hat{\xi}_{(\alpha|\chi|}\triangledown^{\epsilon}\phi_{\beta)\delta\epsilon}\nonumber\\&+20\phi^{\chi}\phi^{\delta}\hat{\xi}_{(\alpha}{}^{\epsilon}\triangledown_{|\chi|}\phi_{\beta)\delta\epsilon}+16\phi_{(\alpha}\hat{\xi}^{\chi\delta}\phi_{\beta)\chi}{}^{\epsilon}\triangledown_{\epsilon}\phi_{\delta}+20\phi^{\chi}\hat{\xi}_{(\alpha}{}^{\delta}\phi_{\beta)\chi\delta}\triangledown^{\epsilon}\phi_{\epsilon}+8\phi^{\chi}\hat{\xi}_{(\alpha}{}^{\delta}\phi_{\beta)\delta}{}^{\epsilon}\triangledown_{\chi}\phi_{\epsilon}\nonumber\\&-8\phi_{(\alpha}\hat{\xi}_{\beta)}{}^{\chi}\phi_{\chi}{}^{\delta\epsilon}\triangledown^{\xi}\phi_{\delta\epsilon\xi}+8\phi_{(\alpha}\hat{\xi}_{\beta)}{}^{\chi}\phi^{\delta\epsilon\xi}\triangledown_{\chi}\phi_{\delta\epsilon\xi}+16\phi_{(\alpha}\hat{\xi}_{\beta)}{}^{\chi}\phi^{\delta\epsilon\xi}\triangledown_{\delta}\phi_{\chi\epsilon\xi}-24\phi_{(\alpha}\hat{\xi}^{\chi\delta}\phi_{\beta)}{}^{\epsilon\xi}\triangledown_{\epsilon}\phi_{\chi\delta\xi}\nonumber\\&-8\phi_{(\alpha}\hat{\xi}^{\chi\delta}\phi_{|\chi}{}^{\epsilon\xi}\triangledown_{\epsilon|}\phi_{\beta)\delta\xi}-8\phi^{\chi}\hat{\xi}_{(\alpha|\chi|}\phi_{\beta)}{}^{\delta\epsilon}\triangledown^{\xi}\phi_{\delta\epsilon\xi}+8\phi^{\chi}\hat{\xi}_{(\alpha}{}^{\delta}\phi_{\beta)}{}^{\epsilon\xi}\triangledown_{\chi}\phi_{\delta\epsilon\xi}+8\phi^{\chi}\hat{\xi}_{(\alpha|\chi|}\phi^{\delta\epsilon\xi}\triangledown_{\beta)}\phi_{\delta\epsilon\xi}\nonumber\\&-20\phi^{\chi}\hat{\xi}_{(\alpha}{}^{\delta}\phi_{|\chi|}{}^{\epsilon\xi}\triangledown_{\beta)}\phi_{\delta\epsilon\xi}-20\phi^{\chi}\hat{\xi}_{(\alpha}{}^{\delta}\phi_{|\chi}{}^{\epsilon\xi}\triangledown_{\delta|}\phi_{\beta)\epsilon\xi}-32\phi^{\chi}\hat{\xi}_{(\alpha}{}^{\delta}\phi_{|\chi}{}^{\epsilon\xi}\triangledown_{\epsilon|}\phi_{\beta)\delta\xi}-16\phi^{\chi}\hat{\xi}_{(\alpha}{}^{\delta}\phi_{|\delta}{}^{\epsilon\xi}\triangledown_{\chi|}\phi_{\beta)\epsilon\xi}\nonumber\\&+8\phi^{\chi}\hat{\xi}_{\chi}{}^{\delta}\phi_{(\alpha}{}^{\epsilon\xi}\triangledown_{|\epsilon|}\phi_{\beta)\delta\xi}-40\phi^{\chi}\hat{\xi}^{\delta\epsilon}\phi_{(\alpha|\delta}{}^{\xi}\triangledown_{\chi|}\phi_{\beta)\epsilon\xi}-12\hat{\xi}_{(\alpha}{}^{\chi}\phi_{\beta)}{}^{\delta\epsilon}\phi_{\chi\delta\epsilon}\triangledown^{\xi}\phi_{\xi}-20\hat{\xi}_{(\alpha}{}^{\chi}\phi_{\beta)}{}^{\delta\epsilon}\phi_{\chi\delta}{}^{\xi}\triangledown_{\epsilon}\phi_{\xi}\nonumber\\&-12\hat{\xi}_{(\alpha}{}^{\chi}\phi_{\beta)}{}^{\delta\epsilon}\phi_{\chi\delta}{}^{\xi}\triangledown_{\xi}\phi_{\epsilon}-8\hat{\xi}_{(\alpha}{}^{\chi}\phi_{\beta)}{}^{\delta\epsilon}\phi_{\delta\epsilon}{}^{\xi}\triangledown_{\chi}\phi_{\xi}-8\hat{\xi}_{(\alpha}{}^{\chi}\phi_{|\chi}{}^{\delta\epsilon}\phi_{\delta\epsilon|}{}^{\xi}\triangledown_{\beta)}\phi_{\xi}\nonumber\\&-16\hat{\xi}_{(\alpha}{}^{\chi}\phi_{|\chi}{}^{\delta\epsilon}\phi_{\delta\epsilon}{}^{\xi}\triangledown_{\xi|}\phi_{\beta)}+8\hat{\xi}^{\chi\delta}\phi_{(\alpha|\chi|}{}^{\epsilon}\phi_{\beta)\delta}{}^{\xi}\triangledown_{\epsilon}\phi_{\xi}-12\hat{\xi}^{\chi\delta}\phi_{(\alpha|\chi|}{}^{\epsilon}\phi_{\beta)\epsilon}{}^{\xi}\triangledown_{\delta}\phi_{\xi}\nonumber\\&-40\hat{\xi}^{\chi\delta}\phi_{(\alpha|\chi|}{}^{\epsilon}\phi_{\beta)\epsilon}{}^{\xi}\triangledown_{\xi}\phi_{\delta}-12\hat{\xi}^{\chi\delta}\phi_{(\alpha|\chi}{}^{\epsilon}\phi_{\delta\epsilon|}{}^{\xi}\triangledown_{\beta)}\phi_{\xi}+12\hat{\xi}^{\chi\delta}\phi_{(\alpha}{}^{\epsilon\xi}\phi_{|\chi\delta\epsilon|}\triangledown_{\beta)}\phi_{\xi}-24\hat{\xi}^{\chi\delta}\phi_{(\alpha|\chi}{}^{\epsilon}\phi_{\delta\epsilon}{}^{\xi}\triangledown_{\xi|}\phi_{\beta)}\nonumber\\&+24\hat{\xi}^{\chi\delta}\phi_{(\alpha}{}^{\epsilon\xi}\phi_{|\chi\delta\epsilon}\triangledown_{\xi|}\phi_{\beta)}-16\hat{\xi}_{(\alpha}{}^{\chi}\phi_{\beta)\chi}{}^{\delta}\phi_{\delta}{}^{\epsilon\xi}\triangledown^{\gamma}\phi_{\epsilon\xi\gamma}+8\hat{\xi}_{(\alpha}{}^{\chi}\phi_{\beta)\chi}{}^{\delta}\phi^{\epsilon\xi\gamma}\triangledown_{\epsilon}\phi_{\delta\xi\gamma}+32\hat{\xi}_{(\alpha}{}^{\chi}\phi_{\beta)}{}^{\delta\epsilon}\phi_{\chi\delta}{}^{\xi}\triangledown^{\gamma}\phi_{\epsilon\xi\gamma}\nonumber\\&-8\hat{\xi}_{(\alpha}{}^{\chi}\phi_{\beta)}{}^{\delta\epsilon}\phi_{\chi}{}^{\xi\gamma}\triangledown_{\xi}\phi_{\delta\epsilon\gamma}+8\hat{\xi}_{(\alpha}{}^{\chi}\phi_{\beta)}{}^{\delta\epsilon}\phi_{\delta\epsilon}{}^{\xi}\triangledown^{\gamma}\phi_{\chi\xi\gamma}-8\hat{\xi}_{(\alpha}{}^{\chi}\phi_{\beta)}{}^{\delta\epsilon}\phi_{\delta}{}^{\xi\gamma}\triangledown_{\chi}\phi_{\epsilon\xi\gamma}\nonumber\\&-8\hat{\xi}_{(\alpha}{}^{\chi}\phi_{|\chi}{}^{\delta\epsilon}\phi_{\delta|}{}^{\xi\gamma}\triangledown_{\beta)}\phi_{\epsilon\xi\gamma}+8\hat{\xi}_{(\alpha}{}^{\chi}\phi_{|\chi}{}^{\delta\epsilon}\phi_{\delta\epsilon|}{}^{\xi}\triangledown^{\gamma}\phi_{\beta)\xi\gamma}+40\hat{\xi}_{(\alpha}{}^{\chi}\phi_{|\chi}{}^{\delta\epsilon}\phi_{\delta}{}^{\xi\gamma}\triangledown_{\xi|}\phi_{\beta)\epsilon\gamma}\nonumber\\&+8\hat{\xi}_{(\alpha}{}^{\chi}\phi^{\delta\epsilon\xi}\phi_{|\delta\epsilon}{}^{\gamma}\triangledown_{\xi|}\phi_{\beta)\chi\gamma}+12\hat{\xi}^{\chi\delta}\phi_{(\alpha|\chi|}{}^{\epsilon}\phi_{\beta)\epsilon}{}^{\xi}\triangledown^{\gamma}\phi_{\delta\xi\gamma}-24\hat{\xi}^{\chi\delta}\phi_{(\alpha|\chi|}{}^{\epsilon}\phi_{\beta)}{}^{\xi\gamma}\triangledown_{\epsilon}\phi_{\delta\xi\gamma}\nonumber\\&+16\hat{\xi}^{\chi\delta}\phi_{(\alpha|\chi|}{}^{\epsilon}\phi_{\beta)}{}^{\xi\gamma}\triangledown_{\xi}\phi_{\delta\epsilon\gamma}+28\hat{\xi}^{\chi\delta}\phi_{(\alpha}{}^{\epsilon\xi}\phi_{\beta)\epsilon}{}^{\gamma}\triangledown_{\xi}\phi_{\chi\delta\gamma}+12\hat{\xi}^{\chi\delta}\phi_{(\alpha|\chi}{}^{\epsilon}\phi_{\delta\epsilon|}{}^{\xi}\triangledown^{\gamma}\phi_{\beta)\xi\gamma}\nonumber\\&+64\hat{\xi}^{\chi\delta}\phi_{(\alpha|\chi}{}^{\epsilon}\phi_{\epsilon}{}^{\xi\gamma}\triangledown_{\xi|}\phi_{\beta)\delta\gamma}-12\hat{\xi}^{\chi\delta}\phi_{(\alpha}{}^{\epsilon\xi}\phi_{|\chi\delta\epsilon|}\triangledown^{\gamma}\phi_{\beta)\xi\gamma}-8\hat{\xi}^{\chi\delta}\phi_{(\alpha}{}^{\epsilon\xi}\phi_{|\chi\delta}{}^{\gamma}\triangledown_{\epsilon|}\phi_{\beta)\xi\gamma}\nonumber\\&
+8\hat{\xi}^{\chi\delta}\phi_{(\alpha}{}^{\epsilon\xi}\phi_{|\chi\delta}{}^{\gamma}\triangledown_{\gamma|}\phi_{\beta)\epsilon\xi}
-4\hat{\xi}^{\chi\delta}\phi_{(\alpha}{}^{\epsilon\xi}\phi_{|\chi\epsilon}{}^{\gamma}\triangledown_{\xi|}\phi_{\beta)\delta\gamma}+4\hat{\xi}^{\chi\delta}\phi_{(\alpha}{}^{\epsilon\xi}\phi_{|\chi\epsilon}{}^{\gamma}\triangledown_{\gamma|}\phi_{\beta)\delta\xi} \nonumber\\&
-16\hat{\xi}^{\chi\delta}\phi_{(\alpha}{}^{\epsilon\xi}\phi_{|\epsilon\xi}{}^{\gamma}\triangledown_{\gamma|}\phi_{\beta)\chi\delta} -8\phi^{\chi}\hat{\xi}^{\delta\epsilon}\phi_{\alpha\beta}{}^{\xi}\triangledown_{\chi}\phi_{\delta\epsilon\xi}\bigr) \ . 
\end{align}


\bibliographystyle{mystyle5}
\bibliography{references}

\end{document}